\documentclass{article}
\usepackage[utf8]{inputenc}
\usepackage{url}
\usepackage{color}
\usepackage{graphicx}
\usepackage{bm, amsmath}
\usepackage{hyperref}
\usepackage{subcaption}
\usepackage{tikz}
\usepackage[margin=1in]{geometry}
\usepackage{xcolor}
\usepackage{fontawesome5} 
\usepackage{adjustbox}
\usepackage{multirow}

\title{A new adaptive two-layer model for opinion spread in hypergraphs: parameter sensitivity and estimation}

\author{Ágnes Backhausz\footnote{Corresponding author. ELTE Eötvös Loránd University, Budapest, Hungary and HUN-REN Alfréd Rényi Institute of Mathematics. Email: \tt{agnes.backhausz@ttk.elte.hu}}, Villő Csiszár\footnote{ELTE Eötvös Loránd University, Budapest, Hungary, Faculty of Science, Department of Probability Theory and Statistics}, \\ Balázs Csegő Kolok\footnote{HUN-REN Alfréd Rényi Institute of Mathematics}, Damján Tárkányi \footnote{ELTE Eötvös Loránd University, Budapest, Hungary, Faculty of Science, Department of Probability Theory and Statistics.}, András Zempléni\footnote{ELTE Eötvös Loránd University, Budapest, Hungary, Faculty of Science, Department of Probability Theory and Statistics.}}

\begin{document}

\maketitle

\begin{abstract} When  opinion spread is studied, peer pressure is often modeled by interactions of more than two individuals (higher-order interactions). In our work, we introduce a two-layer random hypergraph model, in which hyperedges represent households and workplaces. Within this overlapping, adaptive structure, individuals react if their opinion is in majority in their groups. The process  evolves through random steps: individuals can either change their opinion, or quit their workplace and join another one in which their opinion belongs to the majority. Based on computer simulations, our first goal is to describe the effect of the parameters responsible for the probability of changing opinion and quitting workplace on the homophily and speed of polarization. We also analyze the model  as a Markov chain, and study the frequency of the absorbing states. Then, we quantitatively compare how different statistical and machine learning methods, in particular, linear regression, xgboost and a convolutional neural network perform for estimating these probabilities, based on partial information from the process, for example, the distribution of opinion configurations within households and workplaces. Among other observations, we conclude that all methods can achieve the best results under appropriate circumstances, and that the amount of information that is necessary to provide good results depends on the strength of the peer pressure effect.    
\end{abstract}

\noindent{\bf Keywords.} adaptive hypergraph, opinion spread, polarization 

\noindent{\bf MSC classification.} 91D30, 60J20

\section{Introduction}

Random networks have been widely used to model spreading processes, such as epidemics, opinions influencing each other on social networks, or circulation of viruses on the internet \cite{acemoglu2011opinion, berger2005spread, kiss2017mathematics}. It is also often observed that some of these dynamics cannot be understood without including higher-order interactions in our models.  For example, the effect of peer pressure, when individuals are very motivated either to change their opinion or to leave their community if they belong to a small minority within their group, cannot always be modeled only with changes along edges connecting pairs of vertices. Hence, it is natural to study hypergraphs as well: in general, we have a (finite) set of vertices $V$, and a set of hyperedges $\mathcal H$, where each hyperedge $H\in \mathcal H$ is a subset of $V$. Hyperedges can represent groups such as households, workplaces, sports teams, etc.  A random hypergraph can be obtained when we choose $\mathcal H$ randomly. Then, if a vertex $v\in V$ has a certain opinion, the probability of changing it can depend on the configuration of opinions in the hyperedges containing $v$ in a complex way. For example, if $v$ is contained in a hyperedge $H$ of size $5$ such that all the other vertices in $H$ have the opposite opinion from $v$, then $v$ is more likely to change than if it just has $4$ single neighbors with opposite opinion throughout the network from 4 different groups.

Within this framework, various dynamics have been introduced in the literature. For example, Neuh\"auser et al.\ \cite{neuhauser2021consensus, neuhauser2022consensus} study a model in which the evolution of the opinion of vertices (represented by real numbers) is governed by a deterministic system of differential equations, with appropriate interaction functions determining the effect of opinions within each hyperedge. They found that the behavior of the system is significantly different when hyperedges represent interactions occurring one after each other in a given order, compared to the case when we have a static hypergraph with fixed hyperedges.  A possible extension with applications for social dynamics is analysed with the help of computer simulations by Moussa\"id et al.\ \cite{moussaid2013social}. In this case, a confidence parameter is also assigned to each individual. Among other results, the authors found that a minority of $15\%$ is able to achieve a majority for their opinion, if their confidence is high enough; higher-order interactions, effects of groups also play a crucial role in this. For an overview on hypergraphs, higher-order networks and their applications to biology and neuroscience, see e.g.\ the survey of Bick et al.  \cite{bick2023higher}.

In our model, an adaptive hypergraph structure will also play an important role: individuals that are representing the minority opinion in a hyperedge, leave their group and join another one where their opinion is the majority, with a certain probability. Similar dynamics have also been studied in the literature. For example, Horstmeyer and Kuehn \cite{horstmeyer2020adaptive}  studied adaptive models in a slightly different setting (simplicial complexes).  Papanikolaou  et al.\ \cite{papanikolaou2022consensus} studied adaptive voter models on hypergraphs. The expected advantage of the opinions is derived in their paper, together with computer simulations. Golovin, M\"olter and Kuehn \cite{golovin2024polyadic} introduced an adaptive dynamics on hypergraphs, and studied stability and fragmentation.  We remark that such dynamics are not always symmetric in applications: Bermiss and McDonald \cite{bermiss2018ideological} found that leaving the workplace (which will be an important feature in our model) is more likely for conservative individuals than for liberal individuals, if the majority opinion is different from their own choice.

For our study, we introduce a special two-layer model, in which one of the layers (representing households) is deterministic, but the layer representing workplaces consists of randomly chosen hyperedges. That is, the way in which the hyperedges intersect each other has a rich structure. Then, as it is common in the discrete opinion case, the process evolves randomly, with transition rates given by the actual configurations of the hyperedges.  Our model is discrete in the time domain, e.g. potential changes occur only at given equidistant time points. Our approach is symmetric in the sense that we do not differentiate between the roles of the two opinions $\{A, B\}$. On the other hand, to make the model flexible and to be able to observe different behaviors, we will have five different parameters, which govern how easily the vertices change their opinion or leave their workplace when they find themselves in a minority. The dynamics may tend towards homogeneous groups, which often occur in real life as well ("opinion-bubbles" or "polarization"). In addition, as we will see in the next section, each vertex belongs to two groups (one in each layer), and if it represents a minority opinion in both of these groups, that increases further the probability of changing. That is, peer pressure effects are included as interactions between different groups as well. This makes the dynamics significantly different from the graph cases (only using pairwise interactions), which have been studied from the point of view of parameter estimation in our earlier works \cite{backhausz2024estimating, backhausz2024parameter}. 

First we introduce our model in Section \ref{sec:model}. After observing the behavior of the model for a toy example, we base our analysis on large-scale simulations.  
Next we investigate the effect of the different parameters on the spreading process or the speed of polarization (Sections \ref{sec:markov} and  \ref{simu}). Another important goal is to compare different statistical methods that provide good estimates for the unknown parameters (Section \ref{sec:estimation_res}). Our tools applied in this new environment include the XGBoost method and convolutional neural networks as well. We study what kind of information is needed for a good estimate.

\section{The model}\label{sec:model}

In this section, we provide a detailed description of the two-layer random hypergraph model and the spreading process that we analyze.

In our model, the number of vertices (denoted by $n$) is fixed, and the number of hyperedges depends only on $n$. There are two types of hyperedges, called "households" (first layer of the hypergraph) and "workplaces" (second layer), such that every vertex belongs to exactly one household and to exactly one workplace. This hypergraph does not represent perfectly the real structure of social networks (e.g.\ their size will not match the real average sizes); what we had in mind were groups of people who meet and exchange opinions regularly, with large enough intersections to observe non-trivial spreading dynamics. The first layer is a deterministic structure: we divide the $n$ (labelled) vertices into groups of equal size,  for example, vertices with label 1-5, 6-10, etc. form these hyperedges (we assume $n$ to be divisible by $5$ for simplicity). Contrarily, the second layer consists of hyperedges formed from randomly chosen vertices, whose size can change over time. The initial size of workplaces is also 5, to have a balance between the roles of the two layers. We assign the vertices to the hyperedges uniformly at random, independently of the first layer.  

Once we have the underlying hypergraph, first we assign opinion $A$ or $B$ to every vertex, according to a given distribution. During the spreading process (with discrete time-steps), vertices might change their opinion or change workplaces randomly, as follows (notice that households never change, but workplaces and their size might change). 

Our model contains the following parameters corresponding to the dynamics of the process: 
\begin{itemize}
    \item $0 <\beta\leq 1$: opinion change parameter
    \item $0\leq q\leq 1$: workplace change parameter
    \item $0< r_1\leq 1, 0< r_2\leq 1$: thresholds for opinion and workplace change resp.
    \item $0< \lambda\leq 1$: weight of workplaces
\end{itemize}

Then the dynamics is as follows. In each step, an individual $v$ is chosen uniformly at random, and we may assume without loss of generality that his/her opinion is $A$. Let $h$ and $w$ denote the proportion of opinion $A$ in his/her household and workplace, respectively.

\begin{enumerate}
    \item {\bf Changing opinion:} Let us denote  by $a=\frac{h + \lambda \cdot w}{1+\lambda}$ the weighted average of $h$ and $w$. If $a<r_1$, then individual $v$ changes opinion with probability $\beta\cdot (r_1-a)$ .
    
    \item {\bf Changing workplace}: If $v$'s opinion has not changed in  step 1, and $w<r_2$, then with probability $q \cdot (r_2 - w)$ $v$ leaves its workplace and joins a new one. The choice of the new workplace is uniform at random among the workplaces where $v$'s opinion has a majority (that is, the proportion is strictly larger than $0.5$).  If there is no such workplace, the choice is uniform at random among all workplaces.

\end{enumerate}

 The first step corresponds to "peer pressure" within communities, while the second one introduces flexibility in the hypergraph structure, also with a preference for communities where the individuals belong to the majority ("opinion bubbles").  That is, the smaller the proportion of $v$'s opinion is at his workplace, the larger the probability of joining another community where he has the majority. 
We will consider the special case $r_1=r_2=1$ as the {\it linear model}, as in this case conditions $a<r_1$ and $w<r_2$ hold unless $h=w=1$ and all the neighbors of $v$ have the same opinion. Therefore, the probability of changing opinion or workplace is a linear function of $h$ and $w$. On the other hand, if we set $r_1=r_2=0.5$, conditions $a<r_1$ and $w<r_2$ are satisfied only if $v$'s opinion is in the minority, and hence the probability of changing is not a simple linear function of the proportions $h$ and $w$. We will refer to this case as the {\it nonlinear model}.  In this version, peer pressure is stronger in the sense that opinions having a majority cannot be changed (until the majority is preserved). Parameter $\lambda$ is responsible for the weight of workplaces in the opinion-change step; in our model, this had the fixed value $\lambda=0.5$. That is, the layer of households is more influential, but workplaces are also taken into account in the dynamics, with a relatively large weight.

\section{Our model as a Markov chain} \label{sec:markov} 

Opinion spread as we model it is a discrete-time Markov chain on a finite state space $\mathcal{S} \subset \{A, B\}^V\times\{1, \ldots, W\}^V$, where $W$ is the number of workplace hyperedges. Observe that in our model, workplaces are never eliminated, thus $\mathcal{S}$ is only a subset of $\{A, B\}^V\times\{1, \ldots, W\}^V$. Moreover, for 
the chosen $w<r_2 = 0.5$ condition for workplace change, one-element workplaces will never arise either (if in the initial configuration all workplaces are of size at least two).

If all household and workplace hyperedges are homogeneous in the sense that only one opinion is represented in each of them, then the corresponding state of the Markov chain is absorbing. Indeed, since $h=w=1$, neither $a<r_1$ nor $w<r_2$ can be satisfied. On the other hand, this does not imply that only one opinion is represented in the hypergraph. Since the vertices can change their workplace, the hypergraph may split into several connected components, some consisting entirely of opinion A, and others entirely of opinion B.

In the linear model, the states described above are the only absorbing states, since if there is at least one inhomogeneous hyperedge, then $a<1$ for the individuals belonging to that hyperedge, and hence opinion change is possible (here we suppose $\lambda >0$, but the claim is true for $\lambda = 0$ as well). In this case, all other states are transient, since from any nonabsorbing state it is possible to reach an absorbing state by first making all households homogenous (changing the opinions of individuals in nonhomogenous households one by one), then making all workplaces homogenous (by transferring individuals from nonhomogenous workplaces one by one, provided $q>0$). Therefore, starting from any state, the process will eventually reach an absorbing state, with probability $1$.

In the nonlinear model, an absorbing state may have hyperedges with mixed opinions due to the introduction of thresholds $r_1,r_2 <1$ (see the example shown in Figure \ref{fig:smallexample} (a)-(h)). For general values of the parameters, there can exist closed classes consisting of more than one state, in which case the process never "stops." For example, if $r_1$ is close enough to $0$,  then no opinion will ever change, and if in our initial configuration there is only one person with opinion $B$ in the whole population, then this individual will move around between the workplaces forever. Hence we will study the case $r_1=0.5$, when absorbing states appear. Other larger values $r_1$, not very far from $0.5$, would lead to similar dynamics compared to this particular case.

\subsection{Absorbing states for a toy example}

\begin{figure}
    \begin{adjustbox}{center}
        \begin{tikzpicture}[scale=0.8, every node/.style={scale=0.8}]
            \node[circle, draw=gray!60!black, fill=gray!30, minimum size=0.7cm] (job) at (0,0) {};
            \node at (1.5,0) {Workplace};
            \draw[rounded corners, thick] (2.7,-0.3) rectangle (3.7,0.3);
            \node at (4.8,0) {Household};
        \end{tikzpicture}
    \end{adjustbox}
    
    \vspace{0.em}
    \centering
    \begin{subfigure}[t]{0.22\textwidth}
        \centering
        \begin{tikzpicture}[scale=0.35, every node/.style={scale=0.6}]
            \foreach \i in {0,...,9} {
                \node[circle, draw, fill=blue!30, minimum size=0.5cm] (n\i) at (\i,0) {};
            }
            \draw[rounded corners, thick] (-0.5,-0.6) rectangle (4.45,0.6);
            \draw[rounded corners, thick] (4.55,-0.6) rectangle (9.5,0.6);
            \node[circle, draw=gray!60!black, fill=gray!30, minimum size=0.7cm] (w1) at (2,2) {};
            \node[circle, draw=gray!60!black, fill=gray!30, minimum size=0.7cm] (w2) at (7,2) {};
            \draw[-] (n6) to[out=135, in = - 25](w1);
            \draw[-] (n8) to[out=135, in = - 10](w1);
            \foreach \i in {0,2,4} {
                \draw[-] (n\i) -- (w1);
            }
            \draw[-] (n1) to[out=45, in = - 170](w2);
            \draw[-] (n3) to[out=45, in = - 155](w2);
            \foreach \i in {5,7,9} {
                \draw[-] (n\i) -- (w2);
            }

        \end{tikzpicture}
        \caption*{(a)}
    \end{subfigure}
    \hfill
    \begin{subfigure}[t]{0.22\textwidth}
        \centering
        \begin{tikzpicture}[scale=0.35, every node/.style={scale=0.6}]
            \foreach \i in {0,...,4} {
                \node[circle, draw, fill=blue!30, minimum size=0.5cm] (n\i) at (\i,0) {};
            }
            \foreach \i in {5,...,9} {
                \node[circle, draw, fill=red!30, minimum size=0.5cm] (n\i) at (\i,0) {};
            }
            \draw[rounded corners, thick] (-0.5,-0.6) rectangle (4.45,0.6);
            \draw[rounded corners, thick] (4.55,-0.6) rectangle (9.5,0.6);
            \node[circle, draw=gray!60!black, fill=gray!30, minimum size=0.7cm] (w1) at (2,2) {};
            \node[circle, draw=gray!60!black, fill=gray!30, minimum size=0.7cm] (w2) at (7,2) {};
            \foreach \i in {0,...,4} {
                \draw[-] (n\i) -- (w1);
            }
            \foreach \i in {5,...,9} {
                \draw[-] (n\i) -- (w2);
            }
        \end{tikzpicture}
        \caption*{(b)}
    \end{subfigure}
    \hfill
    \begin{subfigure}[t]{0.22\textwidth}
        \centering
        \begin{tikzpicture}[scale=0.35, every node/.style={scale=0.6}]
            \foreach \i in {0,1,2,3,4} {
                \node[circle, draw, fill=blue!30, minimum size=0.5cm] (n\i) at (\i,0) {};
            }
            \foreach \i in {5,6,7,8,9} {
                \node[circle, draw, fill=red!30, minimum size=0.5cm] (n\i) at (\i,0) {};
            }
            \draw[rounded corners, thick] (-0.5,-0.6) rectangle (4.45,0.6);
            \draw[rounded corners, thick] (4.55,-0.6) rectangle (9.5,0.6);
            \node[circle, draw=gray!60!black, fill=gray!30, minimum size=0.7cm] (w1) at (2,2) {};
            \node[circle, draw=gray!60!black, fill=gray!30, minimum size=0.7cm] (w2) at (7,2) {};
            \foreach \i in {0,1,2,5,6,7} {
                \draw[-] (n\i) -- (w1);
            }
            \foreach \i in {3,4,8,9} {
                \draw[-] (n\i) -- (w2);
            }
        \end{tikzpicture}
        \caption*{(c)}
    \end{subfigure}
    \hfill
    \begin{subfigure}[t]{0.22\textwidth}
        \centering
        \begin{tikzpicture}[scale=0.35, every node/.style={scale=0.6}]
            \foreach \i in {0,1,2,3,4} {
                \node[circle, draw, fill=blue!30, minimum size=0.5cm] (n\i) at (\i,0) {};
            }
            \foreach \i in {5,6,7,8,9} {
                \node[circle, draw, fill=red!30, minimum size=0.5cm] (n\i) at (\i,0) {};
            }
            \draw[rounded corners, thick] (-0.5,-0.6) rectangle (4.45,0.6);
            \draw[rounded corners, thick] (4.55,-0.6) rectangle (9.5,0.6);
            \node[circle, draw=gray!60!black, fill=gray!30, minimum size=0.7cm] (w1) at (2,2) {};
            \node[circle, draw=gray!60!black, fill=gray!30, minimum size=0.7cm] (w2) at (7,2) {};
            \foreach \i in {0,1,2,3,5,6,7,8} {
                \draw[-] (n\i) -- (w1);
            }
            \foreach \i in {4,9} {
                \draw[-] (n\i) -- (w2);
            }
        \end{tikzpicture}
        \caption*{(d)}
    \end{subfigure}
    \vspace{0.5em}
    \begin{subfigure}[t]{0.22\textwidth}
        \centering
        \begin{tikzpicture}[scale=0.35, every node/.style={scale=0.6}]
            \foreach \i in {0,1,5,6} {
                \node[circle, draw, fill=blue!30, minimum size=0.5cm] (n\i) at (\i,0) {};
            }
            \foreach \i in {2,3,4,7,8,9} {
                \node[circle, draw, fill=red!30, minimum size=0.5cm] (n\i) at (\i,0) {};
            }
            \draw[rounded corners, thick] (-0.5,-0.6) rectangle (4.45,0.6);
            \draw[rounded corners, thick] (4.55,-0.6) rectangle (9.5,0.6);
            \node[circle, draw=gray!60!black, fill=gray!30, minimum size=0.7cm] (w1) at (2,2) {};
            \node[circle, draw=gray!60!black, fill=gray!30, minimum size=0.7cm] (w2) at (7,2) {};
            \foreach \i in {0,1,5,6} {
                \draw[-] (n\i) -- (w1);
            }
            \foreach \i in {2,3,4,7,8,9} {
                \draw[-] (n\i) -- (w2);
            }
        \end{tikzpicture}
        \caption*{(e)}
    \end{subfigure}
    \hfill
    \begin{subfigure}[t]{0.22\textwidth}
        \centering
        \begin{tikzpicture}[scale=0.35, every node/.style={scale=0.6}]
            \foreach \i in {0,1,2,5,6} {
                \node[circle, draw, fill=blue!30, minimum size=0.5cm] (n\i) at (\i,0) {};
            }
            \foreach \i in {3,4,7,8,9} {
                \node[circle, draw, fill=red!30, minimum size=0.5cm] (n\i) at (\i,0) {};
            }
            \draw[rounded corners, thick] (-0.5,-0.6) rectangle (4.45,0.6);
            \draw[rounded corners, thick] (4.55,-0.6) rectangle (9.5,0.6);
            \node[circle, draw=gray!60!black, fill=gray!30, minimum size=0.7cm] (w1) at (2,2) {};
            \node[circle, draw=gray!60!black, fill=gray!30, minimum size=0.7cm] (w2) at (7,2) {};
            \foreach \i in {0,1,2,5,6} {
                \draw[-] (n\i) -- (w1);
            }
            \foreach \i in {3,4,7,8,9} {
                \draw[-] (n\i) -- (w2);
            }
        \end{tikzpicture}
        \caption*{(f)}
    \end{subfigure}
    \hfill
    \begin{subfigure}[t]{0.22\textwidth}
        \centering
        \begin{tikzpicture}[scale=0.35, every node/.style={scale=0.6}]
            \foreach \i in {0,1,2,3,4,5,6} {
                \node[circle, draw, fill=blue!30, minimum size=0.5cm] (n\i) at (\i,0) {};
            }
            \foreach \i in {7,8,9} {
                \node[circle, draw, fill=red!30, minimum size=0.5cm] (n\i) at (\i,0) {};
            }
            \draw[rounded corners, thick] (-0.5,-0.6) rectangle (4.45,0.6);
            \draw[rounded corners, thick] (4.55,-0.6) rectangle (9.5,0.6);
            \node[circle, draw=gray!60!black, fill=gray!30, minimum size=0.7cm] (w1) at (2,2) {};
            \node[circle, draw=gray!60!black, fill=gray!30, minimum size=0.7cm] (w2) at (7,2) {};
            \foreach \i in {0,1,2,3,4,5,6} {
                \draw[-] (n\i) -- (w1);
            }
            \foreach \i in {7,8,9} {
                \draw[-] (n\i) -- (w2);
            }
        \end{tikzpicture}
        \caption*{(g)}
    \end{subfigure}
    \hfill
    \begin{subfigure}[t]{0.22\textwidth}
        \centering
        \begin{tikzpicture}[scale=0.35, every node/.style={scale=0.6}]
            \foreach \i in {0,1,2,3,4,5,6,7} {
                \node[circle, draw, fill=blue!30, minimum size=0.5cm] (n\i) at (\i,0) {};
            }
            \foreach \i in {8,9} {
                \node[circle, draw, fill=red!30, minimum size=0.5cm] (n\i) at (\i,0) {};
            }
            \draw[rounded corners, thick] (-0.5,-0.6) rectangle (4.45,0.6);
            \draw[rounded corners, thick] (4.55,-0.6) rectangle (9.5,0.6);
            \node[circle, draw=gray!60!black, fill=gray!30, minimum size=0.7cm] (w1) at (2,2) {};
            \node[circle, draw=gray!60!black, fill=gray!30, minimum size=0.7cm] (w2) at (7,2) {};
            \foreach \i in {0,1,2,3,4,5,6,7} {
                \draw[-] (n\i) -- (w1);
            }
            \foreach \i in {8,9} {
                \draw[-] (n\i) -- (w2);
            }
        \end{tikzpicture}
        \caption*{(h)}
    \end{subfigure}
    \caption{Different absorbing states configurations with household and workplace structures in the model for $n=10$. Node color represents opinion; the two colors can be switched respectively to gain homomorphic states. The linear model has only two kinds of absorbing states: (a),(b). The nonlinear model has 8 kinds of absorbing states: (a)-(h).  For example, for state $(e)$, the value $a$ is $0.6$ for blue vertices and $0.73$ for red ones (with $\lambda=0.5$); while in state $(h)$, the value of $a$ is $1$ for vertices in the first household; $0.73$ for blue vertices in the second household, and $0.6$ for the red vertices. Notice that $0.6$ is the smallest value of $a$ which is larger than $0.5$ and can lead to a stable state. State $(a)$ has modifications with an arbitrary workplace structure. Recall that we have strict inequality both in Step 1 and Step 2; otherwise only homogeneous workplaces would remain.} 
    \label{fig:smallexample}
\end{figure}
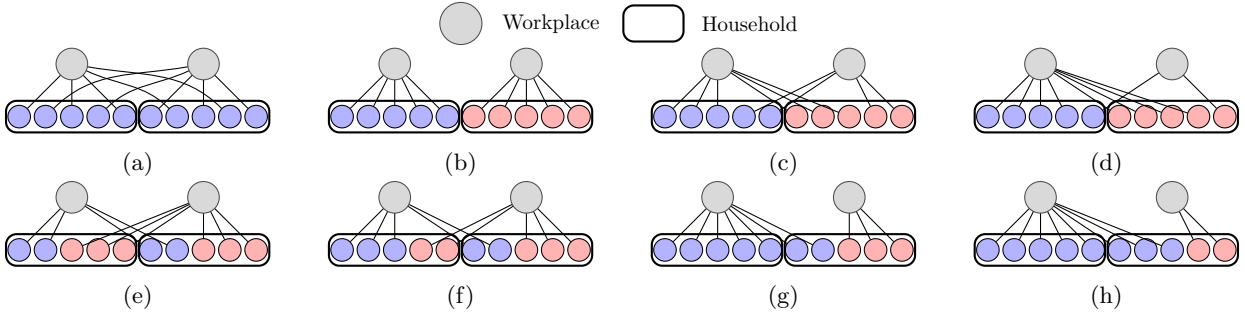

To illustrate the range of possible outcomes, we first consider a small example with $n = 10$ nodes, divided into $2$ households and $2$ workplaces. In this setting, all absorbing states can be enumerated and grouped into structurally equivalent (isomorphic) classes,  which can be obtained from each other by changing all opinions (colors in the figure), these are shown in Figure~\ref{fig:smallexample}. There is no other absorbing state apart from isomorphic states to our list, because all other configurations of opinions and workplaces have a positive probability of transmission to other states of the process.
We start the process from a random initial configuration with 5 nodes with opinion A, chosen randomly, independently of the hyperedges. The probability of ending in each absorbing state class depends on the parameters $\beta$ and $q$. 

\begin{figure}[t]
\centering
         \includegraphics[width=0.8\textwidth]{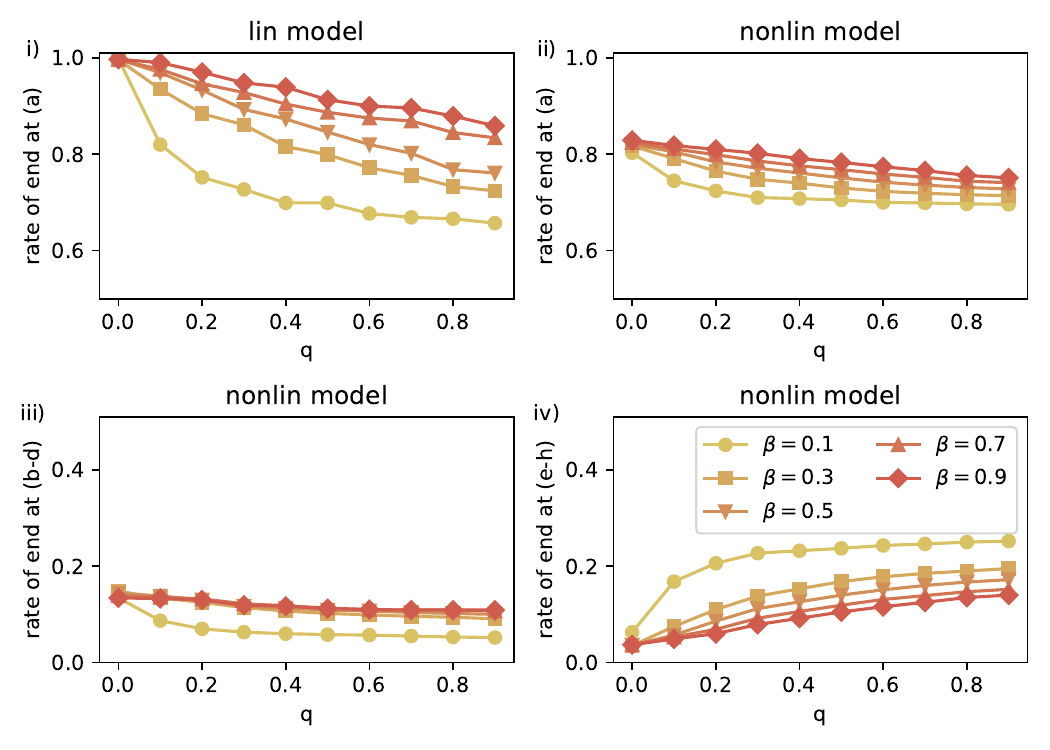}
    \caption{Absorption rates of the states shown in Figure~\ref{fig:smallexample} for the small-scale system with $n = 10$. 
(i) In the linear model, the probability of reaching a fully homogeneous state decreases with increasing $q$ and increases with higher values of $\beta$. 
(ii) In the nonlinear model, the absorption rates show less variation across $q$ and $\beta$, although the same qualitative trend for homogeneous states is observed as in the linear case. 
Absorbing states containing both opinions can be divided into two main categories: (b–d) configurations with two homogeneous households holding opposite opinions, whose absorption rates, shown in (iii), remain nearly constant across the parameter range; and (e–h) configurations with homogeneous workplaces but mixed households, whose absorption rates, shown in (iv), increase with $q$ and decrease with $\beta$.}

    \label{fig:smallexample_prob}
\end{figure}

In the linear model, the system typically requires more iterations to reach absorption than in the nonlinear model. This is because the nonlinear model can terminate in configurations where some hyperedges contain a mix of opinions, whereas the linear model converges only to fully homogeneous hyperedges. As illustrated in Figure~\ref{fig:smallexample_prob}(i)–(ii), the probability of reaching a fully homogeneous state decreases with increasing $q$ in both models: the higher the probability that individuals leave their workplace, the less chance there is for a completely homogeneous final configuration. Furthermore, in the linear model, this dependence is stronger, and the value of $\beta$ also has a significant effect; altogether, we can say that the less stable cases are less likely to lead to homogeneous configurations. In the nonlinear case, we observe the same monotone properties, but in this case, the changes are less significant if we modify any of the parameters. 

The nonlinear model permits the stable coexistence of different opinions within households. Mixed households appear in absorbing states when all workplaces are homogeneous (see Figure~\ref{fig:smallexample}e–h). If an individual belongs to a workplace where everyone shares the same opinion, and their household contains at least one other member with the same opinion, then with $r_1 = r_2 = 0.5$ the individual will neither change opinion nor workplace. In social terms, this suggests that when there is some tolerance  for differing opinions, the system can sustain mixed households while still exhibiting polarization at the workplace level. Figure~\ref{fig:smallexample_prob}(iii)-(iv) presents the probabilities of these mixed states for the nonlinear model; monotonicity with respect to the parameters still holds, but the probabilities of the final configurations are not very sensitive to the parameters: the dynamics seem to be similar, once this kind of stronger peer pressure effect is introduced.

\section{Simulations on a larger scale 
}\label{simu}

\begin{figure}[htb!]
    \centering
    \includegraphics[width=0.8\linewidth]{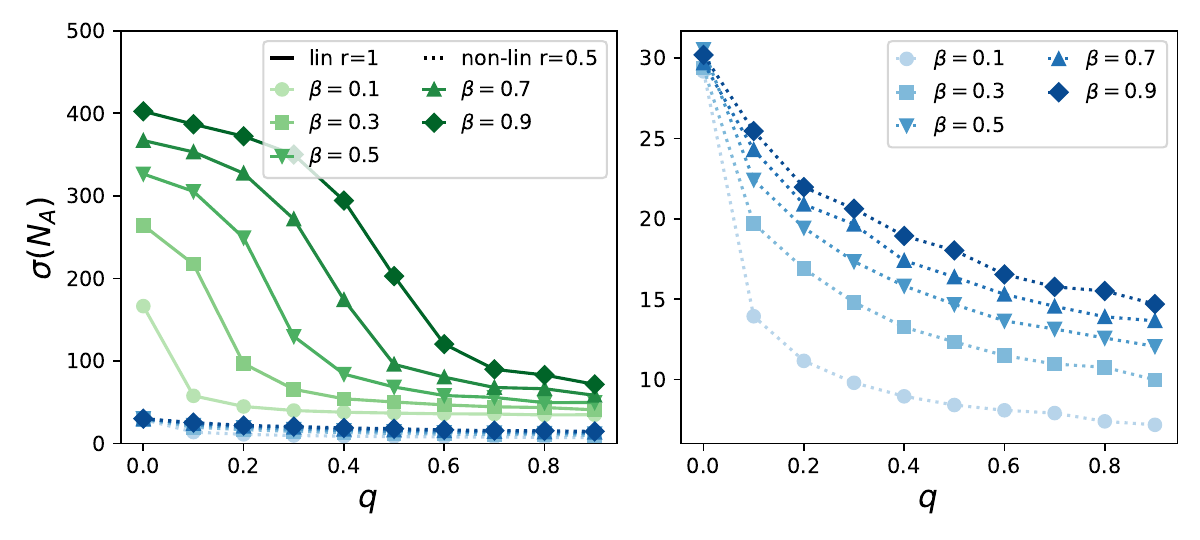}
    \caption{Standard deviation of the number of nodes holding opinion $A$ at the stopping time. Results are based on 500 simulations with $n=1000$, each starting from a balanced initial condition of 500 individuals with opinion $A$ and 500 with opinion $B$, chosen randomly, independently of the hypergraph structure.  Each run lasted $10^6$ steps or stopped in an absorbing state before, so that on average every individual was updated about  1000 times. (a) The linear model (green solid lines) is compared with the nonlinear model (blue dotted lines). (b) Same as (a), but showing only the nonlinear model for clarity. In both models, $\sigma(N_A)$ increases with $\beta$ and decreases with larger $q$. However, the nonlinear model consistently produces smaller deviations than the linear model, highlighting a fundamental difference between the two dynamics—a distinction that will also be reflected in the polarization results.}
    \label{fig:stds}
\end{figure}

Setting $n = 1000$, we performed 500 independent realizations of the process. Since convergence to an absorbing state can take very long, each simulation was terminated in an absorbing state or after reaching a limit of $10^6$ iterations to avoid computational constraints. Firstly, we consider balanced initial conditions with 500 nodes holding opinion $A$ and 500 nodes holding opinion $B$. At the final stopping time, we measure the standard deviation in the number of nodes holding opinion A (Figure~\ref{fig:stds}); notice that the expected numbers of opinions stay constant over time. Overall, the linear model exhibits substantially larger standard deviations than the nonlinear model, which is expected given that the linear dynamics more frequently converge to fully homogeneous states. This behavior is consistent with the small-scale example discussed earlier. Moreover, the deviation tends to increase with higher values of $\beta$, while larger values of $q$ reduce the deviation. In particular, a larger probability of opinion change might increase the overall instability of the model, and since households are fixed, it takes more time for the system to reach homogeneous states; especially for larger $\beta$, Step 2 and workplace change cannot occur that often. On the other hand, if the probability of leaving workplaces is larger, then homogeneous groups can appear very quickly. Workplaces seem to be much more stable, as only vertices having the majority opinion can join them later. Hence, for larger $q$, this has a stabilizing effect for the whole system. 

The mean workplace hyperedge sizes remain $5$ throughout the dynamics. This is a direct consequence of the model construction: the workplace layer is initialized with hyperedges of fixed size, nodes may only rewire between existing workplaces, and no preferential attachment or growth mechanism is present. 

Furthermore, the workplace change rule induces an emergent block-like structure. Since nodes are allowed to move only to workplaces where their opinion is already in the majority, the workplace layer gradually segregates according to opinions. As a result, the long-time structure of the workplace layer resembles a $2$-stochastic block hypergraph, where hyperedges predominantly connect nodes belonging to the same latent group \cite{PhysRevE.110.034312}. In these models, the hyperedge size distribution has finite divergence (and can be approximated by binomial distributions).
In our case, however, this block structure is not imposed a priori but arises dynamically from the adaptive process. This behavior explains why no giant workplaces are observed and hyperedge sizes remain bounded over time, contrasts with evolving hypergraph models that incorporate preferential attachment mechanisms, in which hyperedges may grow unbounded and heavy-tailed hyperedge size distributions can emerge \cite{PhysRevE.106.064310}.

\subsection{Speed of polarization: homophily index and component sizes} 

To investigate the structure of the final states in larger populations, we introduce a measure of homophily in the hypergraph and report the component sizes in the final step of the simulation. In the context of opinion change processes on graphs, the homophily index is defined as the ratio of the number of edges connecting vertices with the same opinion to the total number of edges \cite{NEURIPS2020_58ae23d8}. We generalize this to hypergraphs by defining the homophily index as the ratio of nodes that are part of homogeneous hyperedges \cite{sciadv.abq3200}. We calculate the homophily index separately for household and workplace hyperedges, since they differ from each other, as it can be seen in Figure~\ref{fig:homind}. Together with the statistics about the final component sizes of the hypergraph, we discover an interesting interplay of the parameters $\beta$ and $q$. 

\subsubsection{Linear Model: homogenization and segregation}
In the linear model, the system is driven toward an absorbing state of full polarization. As $t \to \infty$, the probability of reaching a configuration where the hypergraph consists solely of homogeneous hyperedges approaches 1. Our simulations, running for $10^6$ steps, approximate this asymptotic limit. 
As a consequence of that we stop the simulations at a finite time, an interesting phenomenon appears for values of $ q \in [0, 0.4]$: several runs have not yet reached full homogenization, and the ordering of the homophily index curves changes. We see a competitive interplay between opinion change and workplace change, which breaks monotonicity.
As stated before, the homophily index generally converges to 1, indicating that most final states are fully segregated (the two opinions exist only in separate components). However, the path to this state depends on parameter values. As shown in Figure~\ref{fig:homind}a, for low values of $\beta$, the system reaches high homophily rapidly even for small $q$. This occurs because a low $\beta$ increases the rejection rate of opinion changes (Step 1), thereby increasing the frequency of workplace moves (Step 2). These frequent moves accelerate the sorting of agents into homogeneous clusters, inducing a strong peer pressure effect. Conversely, for high $\beta$, opinion change is the dominant mechanism; the system tends to homogenize through consensus within a fixed structure rather than through segregation, leading to a slower rise in structural homophily with respect to $q$.

On Appendix Figure \ref{fig:hh_op_dist} and  \ref{fig:wp_op_dist}, we can also see the polarization of the system for low values of $\beta$ in the distribution of opinions within households and workplaces. Namely,  households consisting entirely of members with opinion $A$ or entirely of members with opinion $B$ are strongly overrepresented compared to mixed households.  We can observe that increasing the workplace change parameter $q$ amplifies this effect in the linear model, leading to higher homophily. We can also study the effect of parameters with Appendix Figure \ref{fig:sumofA_lin}, which shows the trajectories representing the proportion of the individuals with opinion $A$. We see that for lower values of $\beta$ and $q$, there is a significantly larger uncertainty in the evolution of the process; for larger parameters, the proportion stabilizes more quickly. In addition, the behavior of the process depends on both parameters: the effect of changing $\beta$ can be very different for larger $q$. 

The impact of this interplay is most evident in the component size distribution (Appendix Figure~\ref{fig:comp_sizes} and Appendix Figure~\ref{fig:comp_num}). 
\begin{itemize}
    \item \textit{Regime I ($\beta$ dominated):} When $\beta$ is high, and $q$ is low, the network retains a single giant component (size $\approx N$). The system reaches a homogenized consensus (all $A$ or all $B$) without breaking connectivity.
    \item \textit{Regime II ($q$ dominated):} As $q$ increases, the network segregates. The component size distribution shifts from a single value  
    at $N$ to a proper distribution with a mode at $N/2$ (see Appendix Figure~\ref{fig:comp_num}). This indicates that the hypergraph splits into two disjoint, fully polarized components, one for opinion $A$ and one for opinion $B$, effectively creating two separated components.
\end{itemize}
This transition explains the rapid homogenization for low $\beta$: structural splitting isolates conflicting opinions, instantly satisfying the homogeneity condition for all edges within the components.

\subsubsection{Nonlinear Model: mixed coexistence}
\begin{figure}[htb!]
        \centering
        \includegraphics[width=0.8\linewidth]{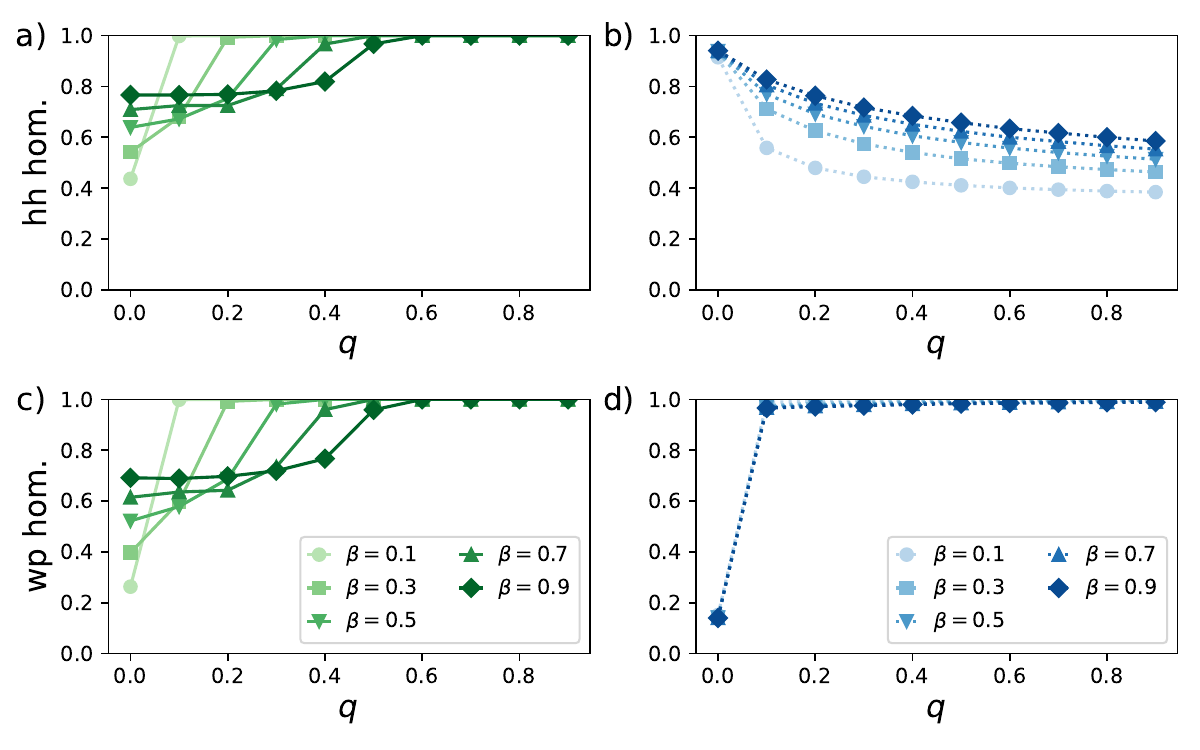}
        \caption{Homophily index for different types of hyperedges: household (top row) and workplaces (bottom row). Here, we used green continuous lines for the linear model and blue dotted lines for the nonlinear model, as before.}
        \label{fig:homind}
\end{figure}

In the nonlinear model, the process does not reach a polarized state with probability $1$. Due to thresholds $r_1$ and $r_2$, there exist absorbing states in which neither opinion nor workplace changes are possible, even though some hyperedges remain mixed (see Figure~\ref{fig:smallexample}). Thus, the nonlinear dynamics allow for the persistence of heterogeneous hyperedges, preventing complete polarization. However, for the workplaces, $q>0$ ensures a complete homogeneity (Figure~\ref{fig:homind} d), and as a larger $q$ allows for a quicker homogenization of the workplaces, the households' homophily will decrease slightly (Figure~\ref{fig:homind} b). For a comparison of the linear and nonlinear model, we refer to Appendix Figure \ref{fig:sumofA_lin} and \ref{fig:sumofA_nonlin}: we can observe that the nonlinear case is much more stable, especially if $\beta$ and $q$ are larger. 

\begin{figure}[htb!]
        \centering
        \includegraphics[width=0.8\linewidth]{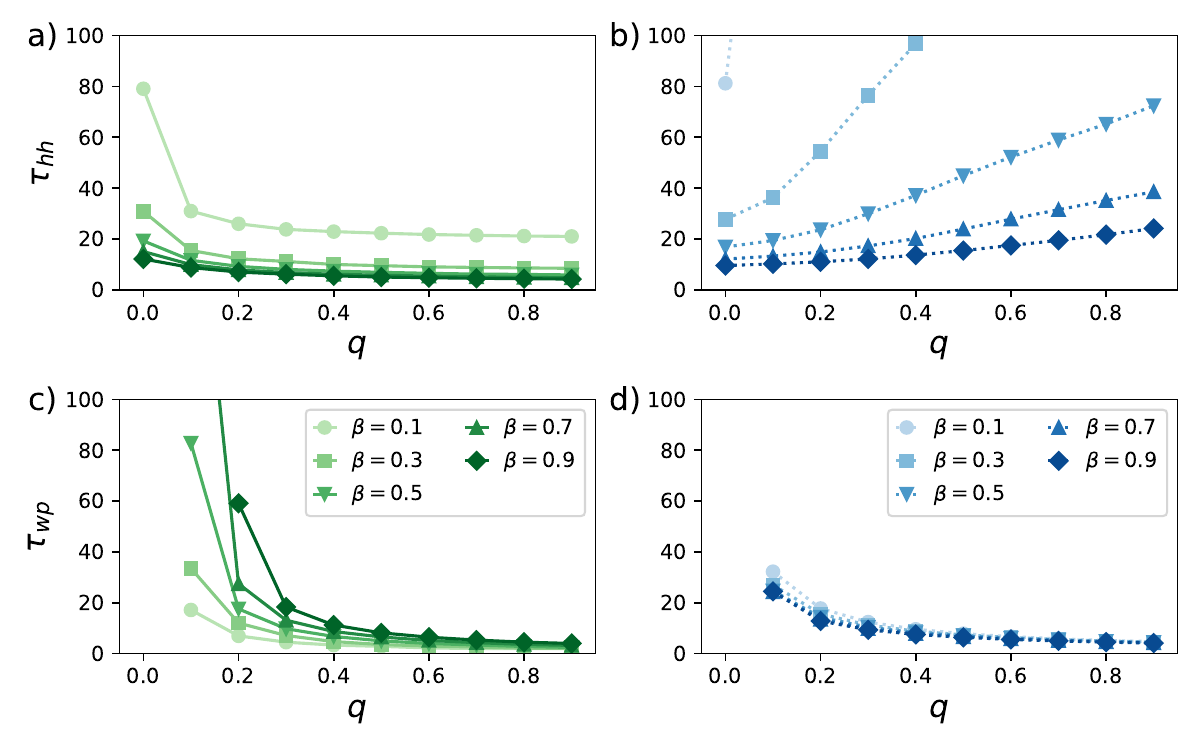}
        \caption{We calculated how long it takes for the process to reach a homophily index above $0.4$. All figures show the average of the first timestep $\tau_{hh},\tau_{wp}$ where the homophily index is greater than $0.4$ amongst the household hyperedges and the workplace hyperedges, respectively. When there is no value shown, that means the homophily did not reach $0.4$ in the first $10^6$ steps, all y-axes are in logging time, so 1 log step means 1000 steps in the model.}
        \label{fig:homtime}
    \end{figure}

Finally, we analyze the speed of polarization by measuring how quickly the homophily index approaches its final value. Specifically, we record the first timestep $\tau_{\rm hh}, \tau_{\rm wp}$ at which the homophily index of households and workplaces, respectively, is above $0.4$ (Figure~\ref{fig:homtime}). This provides an insight into the convergence time toward structural stability in the two layers.

\section{Parameter estimation results}
\label{sec:estimation_res}

We used a large simulated dataset to study the performance of two machine learning methods, XGBoost and CNN, in estimating the parameters $\beta$ and $q$. The setup was essentially the same as in the previous section; the most significant difference is that the number of time steps was only 300, as we are interested in estimates after a shorter observation period as well, and as we needed more detailed information and more repetitions for each choice of parameters. Recall that the population size (number of vertices) was chosen to be $n=1000$, the household size and the initial workplace size were both $5$. The weight of workplaces was fixed as $\lambda=0.5$. We treated the linear ($r_1=r_2=1$) and the nonlinear ($r_1 = r_2 = 0.5$) model separately, that is, we simulated a separate dataset for both cases, as described below. 

The $\beta$ parameter took values in the set
$\{0.1, 0.2, 0.3, 0.4, 0.5, 0.6, 0.7, 0.8, 0.9\}$, while the range of $q$ was $\{0.0, 0.1, 0.2, 0.3, 0.4, 0.5, 0.6, 0.7, 0.8, 0.9\}$. For each combination of the two parameters (i.e. $9\times 10 = 90$ combinations), we simulated the process $500$ times. Each trajectory had length $300$, where one timestep corresponds to $n$ choices of individuals, so in one timestep, every individual has one opportunity to change opinion, on average. For each timestep $t$, we recorded the following statistics:\\
$\bullet$ $N_A(t)$: the number of individuals with opinion $A$ ($1$ dimensional statistic),\\
$\bullet$ $D_{\rm hh}(t)$: the number of households where $k$ people are of opinion $A$ and $5-k$ of opinion $B$, for $k=0, 1, 2, 3, 4, 5$ (i.e. $6$ dimensions),\\ 
$\bullet$ $D_{\rm wp}(t)$: the number of workplaces where the proportion of people with opinion $A$ is between $(k-1)/10$ and $k/10$, $k=1, \ldots, 10$ ($10$ dimensions), \\
$\bullet$ $S_{\rm wp}(t)$: the number of workplaces of size $k$, $k=1, 2, \ldots, 14$ ($14$ dimensions),\\
$\bullet$ $M_{\rm wp}(t)$: the number of movements between workplaces  ($1$ dimensional), \\
$\bullet$ $N_{\rm ch}(t)$: the number of opinion changes  ($1$ dimensional). 
Thus for each time $t$, we recorded the $33$-dimensional statistic $S(t)= (N_A(t), D_{\rm hh}(t), D_{\rm wp}(t), S_{\rm wp}(t), M_{\rm wp}(t), N_{\rm ch}(t))$.  Notice that $N_A(t)$ is a deterministic, linear function of $D_{\rm hh}(t)$: given $t$, the number of individuals with opinion $A$ is the scalar product of $(0,1,2,\ldots, 5)$ and the $6$-dimensional vector $D_{\rm hh}(t)$.

To estimate the parameters, we divided the $500$ trajectories available for all parameter combinations into training and test sets: $400$ trajectories (i.e. $80\%$) were used to train the estimation algorithm, while the remaining $100$ trajectories (i.e. $20\%$) were used to test the precision of the estimate. The error of the estimators was measured by the root mean square error (RMSE).

\subsection{Regression method}\label{reg}

In this simple first approach, at time $t$, the one-dimensional observations till time $t$ and the mean and variance values of $D_{\rm hh}(t), D_{\rm wp}(t)$ and $S_{\rm wp}(t)$ are used as independent variables. Because of the visible nonlinearity, the square of the variances was also included. This means that the number of explanatory variables in the models varies from 20 to 240 for time point $t=20$ and from 300 to 3600 for $t=300$. 
The root mean squared errors for the test set are computed for different amounts of used information. The results are shown in tables \ref{reg1} and \ref{reg2}.

\subsection{XGBoost method}

XGBoost (eXtreme Gradient Boosting) is a powerful algorithm, where weak learners (decision trees) are combined iteratively in order to get a strong method. It is suitable for regression-type problems, like ours -- but it is not especially designed for time series. As it is rather quick for the moderate data sizes we have in mind, we overcome this problem by running the algorithm from scratch for each time point $T$, based on observed data up to time $T$. We use the pre-programmed function in the R package 'xgboost', with booster gbtree (gradient boosted tree).

When running XGBoost in R, most hyperparameters were set at their default values  (eta=0.2, subsample=0.9, nrounds=90, colsample\_bytree=0.8,
min\_child\_weight=0.7). According to our experience, in this case, modification of these hyperparameters does not have a significant effect on the results.

We investigated the properties of the XGBoost estimator, especially its dependence on the availability of information and compared it to a more conventional, regression-based estimator (see Subsection \ref{reg}).
The results are shown in Tables \ref{reg1} and \ref{reg2}.

\subsection{Convolutional Neural Networks for parameter estimation}
In problems of parameter estimation on time series data, a more context-specific neural network architecture than XGBoost is the 1-dimensional Convolutional Neural Network (CNN) architecture \cite{Paul2021-mu, borovykh2017conditional, guessoum2022short}, which is what we chose to employ in this work. Although different other types of artificial neural networks have been used for such tasks - different RNNs such as LSTM and GRUs \cite{zhang2000predicting, jin2019prediction, bai2018empirical} or more recent Transformer architectures \cite{vaswani2017attention, wen2022transformers} among others - the 1d-CNN architecture has limited computational cost compared to Transformers \cite{Tay2022}, while is also shown to outperform recurrent architectures for many sequential modeling tasks \cite{bai2018empirical} and has been used for parameter estimation and forecasting in time series analysis \cite{alizadeh2025epidemic, castro2021stconvs2s, gabrielli2017introducing}. While estimates for propagation-related parameters have been extracted indirectly from forecast data in other studies \cite{cardoso2022modeling, rizvi2022time}, in this paper, we use the simulated dataset to infer the previously introduced parameters directly as output. \\
Our baseline model consists of multiple layers of convolutional kernels with initially gradually increasing, then decreasing number of channels as computation progresses, and finally some fully connected layers as well. Using such a layout is prevalent in similar scenarios to extract and process higher-level features of the data \cite{simonyan2014very, he2015resnet}.\\

\begin{table}[h!]
    \centering
    \begin{tabular}{c|c|c|c|c}
         Layer & $C1D_{i=1\ldots N_{\rm dil}}$ & $MP_{i=1\ldots N_{\rm pool}}$& $C1D_{i=N_{\rm dil}\ldots N_{\rm dil}+N_{\rm con}}$ & $FC_{j=1\ldots N_{\rm fc}}$ \\
         \hline
         Input Channels & $[ \,n_{\rm var}\cdot \left(\frac{n_{\rm max}}{n_{\rm var}} \right)^{\frac{i-1}{N_{\rm dil}}} ] \,$ & $[ \,n_{\rm var}\cdot \left(\frac{n_{\rm max}}{n_{\rm var}} \right)^{\frac{i}{N_{\rm dil}}} ] \,$ & $[ \,n_{\rm max}^{1-\frac{N_{\rm dil}-i-1}{N_{\rm con}}} ] \,$ & 1 \\
         Output Channels & $	[ \,n_{\rm var}\cdot \left(\frac{n_{\rm max}}{n_{\rm var}} \right)^{\frac{i}{N_{\rm dil}}} ] \,$ & $[ \,n_{\rm var}\cdot \left(\frac{n_{\rm max}}{n_{\rm var}} \right)^{\frac{i}{N_{\rm dil}}} ] \,$ & $[ \,n_{\rm max}^{1-\frac{N_{\rm dil}-i}{N_{\rm con}}} ] \,$ & 1 \\
         Input length & $[ \,n_{\rm input}\cdot n_{\rm kernel}^{-i}] \,$ & $[ \,n_{\rm input}\cdot n_{\rm kernel}^{-i} ] \,$ & $[ \,n_{\rm input}\cdot n_{\rm kernel}^{-N_{\rm pool}} ] \,$ & $n_{\rm input} $ \\
    \end{tabular}
    \caption{Architecture of the optimized neural network. The following abbreviations are used: C1D - 1-Dimensional Convolutional layer, MP - Maximum pooling layer, FC - Fully connected layer.\\
    $N_{\rm dil}$, $N_{\rm con}$, $N_{\rm pool}$, $N_{\rm fc}$ are the number of "dilating", "contracting" convolutional layers (i.e. where the number of channels increases and decreases) and the number of maximum pooling layers and the number of fully connected layers respectively.\\
    $n_{\rm var}$, $n_{\rm input}$, $n_{\rm max}$, $n_{\rm kernel}$ are the aggregate dimension of the variables as detailed in Section \ref{sec:estimation_res}, the length of the time series, the maximum number of convolutional channels and the size of the pooling kernel.\\
    After optimization, the following parameters were used: $N_{\rm dil}=8$, $N_{\rm con}=6$, $N_{\rm pool}=4$, $N_{\rm fc}=3$, $n_{\rm max}=1024$ and $n_{\rm kernel}$ is the maximum integer such that the final length after pooling doesn't decrease below $n_{\rm final} = 10$ ($n_{\rm final}$ was the actually optimized parameter).}
    \label{tab:cnn-arch}
\end{table}
The models were fitted to a simulated dataset consisting of 500 samples of time series of 300 simulation timesteps sampled from models with household edgesize $5$ and parameters $r \in \{0.5, 1\}$, and $\beta, q\in [0, 0.9]$ with $0.1$ sampling rate for each combination, i.e. 100000 samples in total. The training of the CNN model involved separating a random 20$\%$ of the dataset for validation. The target variables were the parameters $\beta$ and $q$, respectively. To minimize computational cost, only training and validation sets were used with no separate test set to evaluate performance. Training was carried out for 40 epochs ubiquitously, having chosen a number that yielded the best results on average. \\
The model was implemented using the PyTorch library \cite{paszke2019pytorch}, and the machine learning algorithm was carried out based on the commonplace root mean-squared error (RMSE) as a loss function with the Adam algorithm \cite{kingma2014adam} implemented in the used library.\\
Within this setup, some hyperparameters were optimized, to obtain the best possible results on the dataset, using a random grid search. Optimized parameters related to the learning algorithm were batch size and learning rate, while the number and sizes of the convolutional layers and fully connected layers were the optimizable hyperparameters determining the architecture itself. In addition, to further improve the accuracy of the estimator, we added maximum pooling layers to the architecture, each layer shortening the total length of the processed input, which is initially the time series. With certain configurations of such added layers, the accuracy of the estimator increased significantly, so the defining hyperparameters were also optimized. In the resulting setup, the batch size was 50, learning rate 0.002, while the final dimensions of the layers are shown in Table \ref{tab:cnn-arch}. These choices were then used for all subsequent results.\\
Next, as with the previous model, we examined the contribution of the different types of data in the time series to the quality of the parameter estimation. All subsets of variables were examined for both goal parameters, with results shown in Table \ref{cnn_feat}. Any further estimates were produced using all variables.\\
As with XGBoost, we implemented separately trained CNNs according to the length of the known initial segment of the simulation process. However, since the variance of the CNN estimation error given the same architecture and training setting is much more significant than with XGBoost (see fig. ref), we used the average of 25 independently trained models to indicate performance (resulting in a smoother error curve).

\begin{table}[ht] 
\caption{Overall RMSE(x100) of the CNN-based estimators of $\beta$ and $q$. The first four columns indicate which factors were used. Time=100.} 
\centering 
\label{cnn_feat}
\begin{tabular}{ccccrrrr}
  \hline
$S_{\rm wp}$ & $D_{\rm wp}$ & $D_{\rm hh}$ & $N_A$ & lin. $\beta$ & lin. $q$ & nonlin. $\beta$ & nonlin. $q$\\
  \hline
  0 & 0 & 0 & 1  & 6.07 & 4.51 & 4.29 & 2.99 \\ 
  0 & 0 & 1 & 0  & 6.41 & 4.54 & 4.40 & 2.98 \\ 
  0 & 0 & 1 & 1  & 6.36 & 4.57 & 4.49 & 3.05 \\ 
  0 & 1 & 0 & 1  & 6.57 & 4.80 & 4.57 & 2.88 \\ 
  1 & 0 & 0 & 0  & 6.46 & 4.80 & 4.35 & 2.98 \\ 
  1 & 0 & 0 & 1  & 6.64 & 4.65 & 4.44 & 3.17 \\ 
   \hline
\end{tabular}
\end{table}

\subsection{Results and comparison of the methods}

\begin{figure}
        \centering
        \includegraphics[scale=0.85]{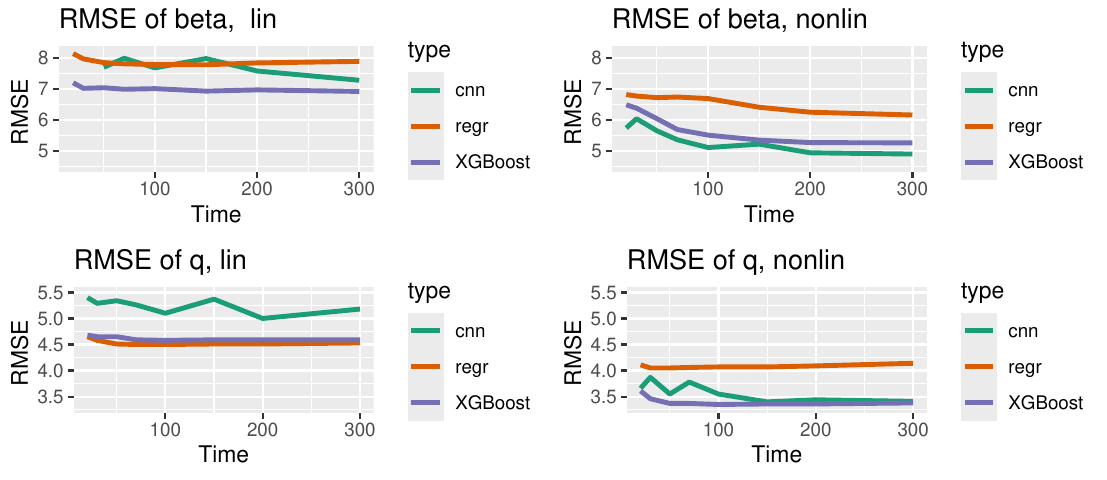}
        \label{fig:rmse_4f_12}
        \caption{ Root mean squared error of the methods, based on $N_A$, $D_{\rm hh}$, $D_{\rm wp}$ and $S_{\rm wp}$, taken from the "no $N_{\rm ch}$ \& $M_{\rm wp}$" lines of tables \ref{tab:all_methods1} and \ref{tab:all_methods2}.}
        \label{fig:regression}
\end{figure}

\begin{table}
\centering
\caption{Overall RMSE(x100) of the estimate of $\beta$  for the investigated hypergraphs. The columns show the number of registrations from the start of the simulation. The first half gives the results for the linear ($r=1$) case, while the second half gives the results for the nonlinear ($r=0.5$) case.  Blue color indicates the convolutional neural networks' results, and red color corresponds to XGBoost. Recall that $N_A$ is a deterministic, linear function of $D_{\rm hh}$. We highlighted the best results in the cases when all three methods were applied with the same training set, and also the second best if its error was almost the same as the error of the best (the difference was at most $3\%$).}
\label{reg1}
\begin{tabular}{l|c|c|c|c|c|c|c|c|} 
method &  20 & 30 & 50 & 70& 100& 150& 200& 300\\
\hline
\multicolumn{9}{c|}{\bf linear model ($r=1$)}   \\ \hline
regr., all information & {\bf 1.52} & {\bf 1.51} & {\bf 1.50} & {\bf 1.50} & {\bf 1.50} & {\bf 1.51} & {\bf 1.52} & {\bf 1.54} \\ 
\hline
\color{blue}{cnn, all information} & 7.95 & 7.41 & 6.71  & 6.58 & 6.44 & 7.37 & 6.43 & 7.34\\
\hline
\color{red}{xgboost, all information} & {\bf 1.55} & {\bf 1.55} & {\bf 1.54} & 1.57 & 1.55 & 1.57 & {\bf 1.53} & {\bf 1.58} \\
\hline
\hline
regr., no $N_{\rm ch}$ & 7.81 & 7.51 & 7.25 & 7.14 & 7.07 & 7.03 & 7.08 & 7.11 \\ 
\hline\hline
regr., no $N_{\rm ch}$ \& $M_{\rm wp}$  & 8.14 & 7.97 & 7.85 & 7.81 & 7.79 & 7.78 & 7.84 & 7.89 \\ 
\hline 
\color{blue}{cnn, no $N_{\rm ch}$ \& $M_{\rm wp}$}& 7.98 & 8.27 & 7.70 & 7.99 & 7.68 & 7.98 & 7.58 & 7.28 \\ 
\hline
\color{red}{xgboost, no $N_{\rm ch}$ \& $M_{\rm wp}$} & {\bf 7.20} & {\bf 7.02} & {\bf 7.04} & {\bf 6.99} & {\bf 7.01} & {\bf 6.93} & {\bf 6.97} & {\bf 6.92} \\
\hline \hline
regr., $D_{\rm hh}$ \& $D_{\rm wp}$ & 8.19 & 8.06 & 7.99 & 7.97 & 7.95 & 7.94 & 7.99 & 8.05 \\ 
\hline
regr., $D_{\rm hh}$  & 8.76 & 8.57 & 8.50 & 8.49 & 8.47 & 8.46 & 8.49 & 8.50 \\ 
\hline
regr., $N_A$ & 25.83 & 25.83 & 25.84 & 25.84 & 25.86 & 25.89 & 25.91 & 25.95 \\ 
\hline
\hline
\multicolumn{9}{c|}{\bf nonlinear model ($r=0.5$)}  \\ \hline 
regr., all information & 6.42 & 6.32 & 6.27 & 6.29 & 6.25 & 6.04 & 5.88 & 5.78 \\ 
\hline
\color{blue}{cnn, all information} & {\bf 6.17} & {\bf 5.54} & {\bf 5.54} & {\bf 5.35} & {\bf 5.32} & {\bf 5.05} & {\bf 5.14} & {\bf 4.82}\\
\hline 
\color{red}{xgboost, all information} & 6.37 & 6.28 & 5.99  & 5.62 & 5.37 & 5.25 & 5.18 & 5.19 \\
\hline \hline
regr., no $N_{\rm ch}$ & 6.72 & 6.59 & 6.51 & 6.53 & 6.51 & 6.30 & 6.13 & 6.00 \\ 
\hline\hline
regr., no $N_{\rm ch}$ \& $M_{\rm wp}$ & 6.82 & 6.77 & 6.72 & 6.74 & 6.69 & 6.41 & 6.25 & 6.16 \\ 
\hline
\color{blue}{cnn, no $N_{\rm ch}$ \& $M_{\rm wp}$}& {\bf 5.74} & {\bf 6.04} & {\bf 5.65} & {\bf 5.36} & {\bf 5.11} & {\bf 5.22} & {\bf 4.94} & {\bf 4.90} \\ 
\hline
\color{red}{xgboost, no $N_{\rm ch}$ \& $M_{\rm wp}$}& 6.49 & 6.38 &6.04 & 5.69 & 5.51 & 5.35 & 5.27 & 5.26\\
\hline \hline
regr., $D_{\rm hh}$ \& $D_{\rm wp}$  & 6.84 & 6.79 & 6.74 & 6.75 & 6.69 & 6.40 & 6.24 & 6.15 \\ 
\hline
regr., $D_{\rm hh}$    & 8.08 & 8.03 & 8.05 & 8.03 & 7.91 & 7.63 & 7.44 & 7.13 \\ 
\hline
regr., $N_A$ & 25.88 & 25.87 & 25.87 & 25.89 & 25.91 & 25.95 & 25.95 & 26.01 \\ 
 \hline
 \hline
 
\end{tabular}
\label{tab:all_methods1}
\end{table}

\begin{table}\centering
\caption{Overall RMSE(x100) of the estimate of $q$ for the used method, for the investigated hypergraphs. The columns show the number of registrations from the start of the simulation. The first half gives the results for the linear ($r=1$) case, while the second half gives the results for the nonlinear ($r=0.5$) case.  }
\label{reg2}
\begin{tabular}{l|c|c|c|c|c|c|c|c|}\hline
method &  20 & 30 & 50 & 70& 100& 150& 200& 300\\
\hline 
 \multicolumn{9}{c|}{\bf linear model ($r=1$)}  \\ \hline
regr., all information & 2.40 & 2.38 & 2.33 & 2.30 & {\bf 2.30} & 2.31 & 2.31 &  2.34 \\ 
\hline
\color{blue}{cnn, all information} & 2.73 & 3.14 & 3.08 & 3.00 & 2.67 & 2.87 &  2.99 & 2.89 \\
 \hline
\color{red}{xgboost, all information} &{\bf 2.18} & {\bf 2.23} & {\bf 2.19} & {\bf 2.17} & {\bf 2.24} & {\bf 2.20} & {\bf 2.22} & {\bf 2.22}  \\
 \hline \hline
regr., no $N_{\rm ch}$ & 2.57 & 2.54 & 2.52 & 2.51 & 2.51 & 2.52 & 2.53 & 2.55 \\ 
 \hline \hline
regr., no $N_{\rm ch}$ \& $M_{\rm wp}$ & {\bf 4.65} & {\bf 4.58} & {\bf 4.51} & {\bf 4.50} & {\bf 4.50} & {\bf 4.51} & {\bf 4.51} & {\bf 4.53} \\ 
 \hline
 \color{blue}{cnn, no $N_{\rm ch}$ \& $M_{\rm wp}$}& 5.40 & 5.29 & 5.34 & 5.26 & 5.10 & 5.37 & 5.00 & 5.18
\\ 
\hline
 \color{red}{xgboost, no $N_{\rm ch}$ \& $M_{\rm wp}$}& {\bf 4.68} & {\bf 4.65} & 4.65 & {\bf 4.59} & {\bf 4.58} & {\bf 4.59} & {\bf 4.59} & {\bf 4.59} \\
\hline \hline
regr.,  $D_{\rm hh}$ \& $D_{\rm wp}$  & 5.60 & 5.45 & 5.32 & 5.30 & 5.31 & 5.31 & 5.31 & 5.33 \\  \hline
regr., $D_{\rm hh}$ & 14.34 & 12.90 & 11.85 & 11.54 & 11.46 & 11.44 & 11.40 & 11.33 \\  \hline
regr., $N_A$ & 28.74 & 28.74 & 28.76 & 28.75 & 28.77 & 28.81 & 28.82 & 28.85 \\ 
   \hline
 \hline 
 \multicolumn{9}{c|}{\bf nonlinear model ($r=0.5$)}\\ \hline 
regr., all information & 3.76 & 3.73 & 3.72 & 3.73 & 3.74 & 3.75 & 3.77 & 3.82 \\ 
\hline
\color{blue}{cnn, all information} & {\bf 2.92} & {\bf 2.82} & {\bf 2.81} & {\bf 2.70} & {\bf 2.68} & {\bf 2.65} & {\bf 2.74} & {\bf 2.77} \\
 \hline
 \color{red}{xgboost, all information} & 3.58 & 3.35 & 3.36 & 3.28 & 3.30 & 3.30 & 3.33 & 3.34 \\
 \hline \hline
regr., no $N_{\rm ch}$  & 3.77 & 3.74 & 3.73 & 3.74 & 3.75 & 3.76 & 3.77 & 3.82 \\  \hline \hline
regr., no $N_{\rm ch}$ \& $M_{\rm wp}$ & 4.11 & 4.05 & 4.05 & 4.06 & 4.07 & 4.07 & 4.09 & 4.14 \\  \hline
\color{blue}{cnn, no $N_{\rm ch}$ \& $M_{\rm wp}$}& {\bf 3.66} & 3.87 & 3.55 & 3.78 & 3.55 & {\bf 3.40} & {\bf 3.44} & {\bf 3.41}  \\ 
\hline
 \color{red}{xgboost, no $N_{\rm ch}$ \& $M_{\rm wp}$} & {\bf 3.61} & {\bf 3.46} & {\bf 3.37} & {\bf 3.37} & {\bf 3.35} & {\bf 3.36} & {\bf 3.36} & {\bf 3.38} \\
\hline \hline
regr., $D_{\rm hh}$ \& $D_{\rm wp}$  & 4.19 & 4.14 & 4.14 & 4.15 & 4.16 & 4.16 & 4.17 & 4.21 \\  \hline
regr., $D_{\rm hh}$ & 15.62 & 14.22 & 14.01 & 13.97 & 13.92 & 13.80 & 13.63 & 13.08 \\  \hline 
regr., $N_A$ & 28.69 & 28.71 & 28.80 & 28.82 & 28.85 & 28.88 & 28.92 & 29.00 \\ 
   \hline
\hline
\end{tabular}
\label{tab:all_methods2}
\end{table}

   In Tables \ref{reg1} and \ref{reg2}, detailed results are presented for the regression method, and the other, more sophisticated methods for the most interesting cases. In addition, Figure \ref{fig:regression} shows some of the results for better visibility. 

If only household information ($N_A$ and $D_{\rm hh}$) are available, the results for $q$ (Table \ref{reg2}) are not surprisingly quite bad. On the other hand, for $\beta$, household distribution is relevant, and the results do not improve much using the workplace information.

It is interesting that while XGBoost is in most of the cases better than the regression method when estimating $q$, this difference is not too large - moreover, for $\beta$ the regression method quite often outperforms XGBoost. In general, the nonlinear case turns out to be the easier one, except when all information is available, in accordance with the remarks about its higher variability in Subsection \ref{simu}, but the number of opinion changes ensures a very good regression and XGBoost estimator for $\beta$ in the linear case. Interestingly, the neural network is substantially better than the other methods only if $\beta$ is estimated in the nonlinear case.

We can also observe that information on opinion change and workplace move significantly improves the quality of the estimates, especially for the estimate of $\beta$ in the linear case; here this information seems to be crucial to get the best results. In the nonlinear case, when peer pressure is stronger, the differences are smaller. Surprisingly,  the convolutional neural network sometimes provides worse results if these additional statistics are available. It also turns out that the errors do not decrease significantly as time goes on and we get more and more information; in certain cases, they even increase over time.

\section{Discussion}
In this paper, we have introduced a novel, adaptive two-layer model for opinion-spreading dynamics. It is based on an underlying hypergraph, for which the workplace hyperedge may change, in order to reduce minority opinions.

We have investigated the process mostly via simulations, but the Markovian structure allowed us to determine the possible absorbing states for a toy example. For larger-scale hypergraphs, simulations have shown the effect of the parameters on the homophily. We have investigated several relevant quantities, like the number and time series of opinion changes and the number of components, mostly by plots, shown in the Appendix. These preliminary results show some interesting observations about the interplay of the parameters corresponding to opinion change probability and workplace change probability, which we plan to investigate in more detail in the future.

The parameters of the models were in the focus of our investigations, especially as our aim was the estimation of these values, based on different methods. It turned out that in the linear model, simple linear regression often performs as well as more sophisticated approaches. In the nonlinear case, however, convolutional neural networks proved to be the best method in most of the cases, while in the linear case, when all information is available, XGBoost is the best.  We also examined which are the most important statistics from the process to get reasonable estimates on the parameter: in the linear setting, with weaker peer pressure, individual-based statistics (number of opinion changes and workplace changes) were crucial; in the nonlinear case, hyperedge-based statistics were sufficient to get good results. 

\medskip
{\bf Acknowledgement.} This work was supported by the National Research, Development and Innovation Office within the framework of the Thematic Excellence Program 2021 -- National Research Subprogramme: “Artificial intelligence, large networks, data security: mathematical foundation and applications” (grant number: TKP2021-NKTA-62). The authors are grateful for Edit Bogn\'ar for useful discussions on the model and the measurement of polarization. 

\bibliography{hipergraf}

@article{kiss2017mathematics,
  title={Mathematics of epidemics on networks},
  author={Kiss, Istv{\'a}n Z and Miller, Joel C and Simon, P{\'e}ter L and others},
  journal={Cham: Springer},
  volume={598},
  number={2017},
  pages={31},
  year={2017},
  publisher={Springer}
}

@article{golovin2024polyadic,
  title={Polyadic opinion formation: the adaptive voter model on a hypergraph},
  author={Golovin, Anastasia and M{\"o}lter, Jan and Kuehn, Christian},
  journal={Annalen der Physik},
  volume={536},
  number={7},
  pages={2300342},
  year={2024},
  publisher={Wiley Online Library}
}

@article{horstmeyer2020adaptive,
  title={Adaptive voter model on simplicial complexes},
  author={Horstmeyer, Leonhard and Kuehn, Christian},
  journal={Physical Review E},
  volume={101},
  number={2},
  pages={022305},
  year={2020},
  publisher={APS}
}

@inproceedings{berger2005spread,
  title={On the spread of viruses on the internet},
  author={Berger, Noam and Borgs, Christian and Chayes, Jennifer and Saberi, Amin},
  booktitle={Proceedings of the 16th ACM-SIAM Symposium on Discrete Algorithm (SODA)},
  pages={301--310},
  year={2005}
}

@article{acemoglu2011opinion,
  title={Opinion dynamics and learning in social networks},
  author={Acemoglu, Daron and Ozdaglar, Asuman},
  journal={Dynamic Games and Applications},
  volume={1},
  pages={3--49},
  year={2011},
  publisher={Springer}
}

@incollection{neuhauser2022consensus,
  title={Consensus dynamics and opinion formation on hypergraphs},
  author={Neuh{\"a}user, Leonie and Lambiotte, Renaud and Schaub, Michael T},
  booktitle={Higher-Order Systems},
  pages={347--376},
  year={2022},
  publisher={Springer}
}

@article{neuhauser2021consensus,
  title={Consensus dynamics on temporal hypergraphs},
  author={Neuh{\"a}user, Leonie and Lambiotte, Renaud and Schaub, Michael T},
  journal={Physical Review E},
  volume={104},
  number={6},
  pages={064305},
  year={2021},
  publisher={APS}
}

@article{backhausz2024estimating,
  title={Estimating the parameters of epidemic spread on two-layer random graphs: a classical and a neural network approach},
  author={Backhausz, {\'A}gnes and Bogn{\'a}r, Edit and Csisz{\'a}r, Vill{\H{o}} and T{\'a}rk{\'a}nyi, Damj{\'a}n and Zempl{\'e}ni, Andr{\'a}s},
  journal={Journal of Statistical Theory and Practice},
  volume={18},
  number={4},
  pages={50},
  year={2024},
  publisher={Springer}
}

@article{backhausz2024parameter,
  title={Parameter estimation of epidemic spread in two-layer random graphs by classical and machine learning methods},
  author={Backhausz, {\'A}gnes and Bogn{\'a}r, Edit and Csisz{\'a}r, Vill{\H{o}} and T{\'a}rk{\'a}nyi, Damj{\'a}n and Zempl{\'e}ni, Andr{\'a}s},
  journal={arXiv preprint arXiv:2407.07118},
  year={2024}
}

@article{paszke2019pytorch,
  title={Pytorch: An imperative style, high-performance deep learning library},
  author={Paszke, Adam and Gross, Sam and Massa, Francisco and Lerer, Adam and Bradbury, James and Chanan, Gregory and Killeen, Trevor and Lin, Zeming and Gimelshein, Natalia and Antiga, Luca and others},
  journal={Advances in neural information processing systems},
  volume={32},
  year={2019}
}

@article{Paul2021-mu, author={S. K. Paul, S. Jana, P. Bhaumik}, year={ (2021)}, title={A multivariate spatiotemporal model of {COVID-19} epidemic using
               ensemble of {ConvLSTM} networks}, journal={Journal of the Institution of Engineers (India) Series B}, volume={102}, issue={6}, pages={137-1142}}

@article{borovykh2017conditional,
  title={Conditional time series forecasting with convolutional neural networks},
  author={Borovykh, Anastasia and Bohte, Sander and Oosterlee, Cornelis W},
  journal={arXiv preprint arXiv:1703.04691},
  year={2017}
}

@article{guessoum2022short,
  title={The short-term prediction of length of day using 1D convolutional neural networks (1D CNN)},
  author={Guessoum, Sonia and Belda, Santiago and Ferrandiz, Jose M and Modiri, Sadegh and Raut, Shrishail and Dhar, Sujata and Heinkelmann, Robert and Schuh, Harald},
  journal={Sensors},
  volume={22},
  number={23},
  pages={9517},
  year={2022},
  publisher={MDPI}
}

@article{zhang2000predicting,
  title={Predicting chaotic time series using recurrent neural network},
  author={Zhang, Jia-Shu and Xiao, Xian-Ci},
  journal={Chinese Physics Letters},
  volume={17},
  number={2},
  pages={88},
  year={2000},
  publisher={IOP Publishing}
}

@inproceedings{jin2019prediction,
  title={Prediction for Time Series with CNN and LSTM},
  author={Jin, Xuebo and Yu, Xinghong and Wang, Xiaoyi and Bai, Yuting and Su, Tingli and Kong, Jianlei},
  booktitle={Proceedings of the 11th international conference on modelling, identification and control (ICMIC2019)},
  pages={631--641},
  year={2019},
  organization={Springer}
}

@article{vaswani2017attention,
  title={Attention is all you need},
  author={Vaswani, Ashish and Shazeer, Noam and Parmar, Niki and Uszkoreit, Jakob and Jones, Llion and Gomez, Aidan N and Kaiser, {\L}ukasz and Polosukhin, Illia},
  journal={Advances in neural information processing systems},
  volume={30},
  year={2017}
}

@article{wen2022transformers,
  title={Transformers in time series: A survey},
  author={Wen, Qingsong and Zhou, Tian and Zhang, Chaoli and Chen, Weiqi and Ma, Ziqing and Yan, Junchi and Sun, Liang},
  journal={arXiv preprint arXiv:2202.07125},
  year={2022}
}

@article{cardoso2022modeling,
  title={Modeling the geospatial evolution of COVID-19 using spatio-temporal convolutional sequence-to-sequence neural networks},
  author={Cardoso, M{\'a}rio and Cavalheiro, Andr{\'e} and Borges, Alexandre and Duarte, Ana Filipa and Soares, Am{\'\i}lcar and Pereira, Maria Jo{\~a}o and Nunes, Nuno Jardim and Azevedo, Leonardo and Oliveira, Arlindo},
  journal={ACM Transactions on Spatial Algorithms and Systems},
  volume={8},
  number={4},
  pages={1--19},
  year={2022},
  publisher={ACM New York, NY}
}

@article{alizadeh2025epidemic,
  title={Epidemic Forecasting with a Hybrid Deep Learning Method Using CNN-LSTM With WOA-GWO Parameter Optimization: Global COVID-19 Case Study},
  author={Alizadeh, Mousa and Samaei, Mohammad Hossein and Seilsepour, Azam and Beheshti, Mohammad TH},
  journal={arXiv preprint arXiv:2503.12813},
  year={2025}
}

@article{Tay2022,
  title = {Efficient Transformers: A Survey},
  volume = {55},
  ISSN = {1557-7341},
  url = {http://dx.doi.org/10.1145/3530811},
  DOI = {10.1145/3530811},
  number = {6},
  journal = {ACM Computing Surveys},
  publisher = {Association for Computing Machinery (ACM)},
  author = {Tay,  Yi and Dehghani,  Mostafa and Bahri,  Dara and Metzler,  Donald},
  year = {2022},
  month = dec,
  pages = {1–28}
}

@article{bai2018empirical,
  title={An empirical evaluation of generic convolutional and recurrent networks for sequence modeling},
  author={Bai, Shaojie and Kolter, J Zico and Koltun, Vladlen},
  journal={arXiv preprint arXiv:1803.01271},
  year={2018}
}

@article{castro2021stconvs2s,
  title={Stconvs2s: Spatiotemporal convolutional sequence to sequence network for weather forecasting},
  author={Castro, Rafaela and Souto, Yania M and Ogasawara, Eduardo and Porto, Fabio and Bezerra, Eduardo},
  journal={Neurocomputing},
  volume={426},
  pages={285--298},
  year={2021},
  publisher={Elsevier}
}

@inproceedings{gabrielli2017introducing,
  title={Introducing deep machine learning for parameter estimation in physical modelling},
  author={Gabrielli, Leonardo and Tomassetti, Stefano and Squartini, Stefano and Zinato, Carlo and others},
  booktitle={Proceedings of the 20th international conference on digital audio effects},
  year={2017}
}

@article{rizvi2022time,
  title={Time series deep learning for robust steady-state load parameter estimation using 1D-CNN},
  author={Rizvi, Syed M Hur},
  journal={Arabian Journal for Science and Engineering},
  volume={47},
  number={3},
  pages={2731--2744},
  year={2022},
  publisher={Springer}
}

@misc{simonyan2014very,
  doi = {10.48550/ARXIV.1409.1556},
  url = {https://arxiv.org/abs/1409.1556},
  author = {Simonyan,  Karen and Zisserman,  Andrew},
  title = {Very Deep Convolutional Networks for Large-Scale Image Recognition},
  publisher = {arXiv},
  year = {2014},
  copyright = {arXiv.org perpetual,  non-exclusive license}
}

@misc{he2015resnet,
  doi = {10.48550/ARXIV.1512.03385},
  url = {https://arxiv.org/abs/1512.03385},
  author = {He,  Kaiming and Zhang,  Xiangyu and Ren,  Shaoqing and Sun,  Jian},
  title = {Deep Residual Learning for Image Recognition},
  publisher = {arXiv},
  year = {2015},
  copyright = {arXiv.org perpetual,  non-exclusive license}
}

@misc{kingma2014adam,
  doi = {10.48550/ARXIV.1412.6980},
  url = {https://arxiv.org/abs/1412.6980},
  author = {Kingma,  Diederik P. and Ba,  Jimmy},
  title = {Adam: A Method for Stochastic Optimization},
  publisher = {arXiv},
  year = {2014},
  copyright = {arXiv.org perpetual,  non-exclusive license}
}

@article{sciadv.abq3200,
author = {Nate Veldt  and Austin R. Benson  and Jon Kleinberg },
title = {Combinatorial characterizations and impossibilities for higher-order homophily},
journal = {Science Advances},
volume = {9},
number = {1},
pages = {eabq3200},
year = {2023},
doi = {10.1126/sciadv.abq3200},
URL = {https://www.science.org/doi/abs/10.1126/sciadv.abq3200},
eprint = {https://www.science.org/doi/pdf/10.1126/sciadv.abq3200},
}

@article{bick2023higher,
  title={What are higher-order networks?},
  author={Bick, Christian and Gross, Elizabeth and Harrington, Heather A and Schaub, Michael T},
  journal={SIAM review},
  volume={65},
  number={3},
  pages={686--731},
  year={2023},
  publisher={SIAM}
}

@article{moussaid2013social,
  title={Social influence and the collective dynamics of opinion formation},
  author={Moussa{\"\i}d, Mehdi and K{\"a}mmer, Juliane E and Analytis, Pantelis P and Neth, Hansj{\"o}rg},
  journal={PloS one},
  volume={8},
  number={11},
  pages={e78433},
  year={2013},
  publisher={Public Library of Science San Francisco, USA}
}

@article{bermiss2018ideological,
  title={Ideological misfit? Political affiliation and employee departure in the private-equity industry},
  author={Bermiss, Y Sekou and McDonald, Rory},
  journal={Academy of Management Journal},
  volume={61},
  number={6},
  pages={2182--2209},
  year={2018},
  publisher={Academy of Management Briarcliff Manor, NY}
}

@article{papanikolaou2022consensus,
  title={Consensus from group interactions: An adaptive voter model on hypergraphs},
  author={Papanikolaou, Nikos and Vaccario, Giacomo and Hormann, Erik and Lambiotte, Renaud and Schweitzer, Frank},
  journal={Physical Review E},
  volume={105},
  number={5},
  pages={054307},
  year={2022},
  publisher={APS}
}

@inproceedings{NEURIPS2020_58ae23d8,
 author = {Zhu, Jiong and Yan, Yujun and Zhao, Lingxiao and Heimann, Mark and Akoglu, Leman and Koutra, Danai},
 booktitle = {Advances in Neural Information Processing Systems},
 editor = {H. Larochelle and M. Ranzato and R. Hadsell and M.F. Balcan and H. Lin},
 pages = {7793--7804},
 publisher = {Curran Associates, Inc.},
 title = {Beyond Homophily in Graph Neural Networks: Current Limitations and Effective Designs},
 url = {https://proceedings.neurips.cc/paper_files/paper/2020/file/58ae23d878a47004366189884c2f8440-Paper.pdf},
 volume = {33},
 year = {2020}
}

@article{PhysRevE.110.034312,
  title = {Stochastic block hypergraph model},
  author = {Pister, Alexis and Barthelemy, Marc},
  journal = {Phys. Rev. E},
  volume = {110},
  issue = {3},
  pages = {034312},
  numpages = {10},
  year = {2024},
  month = {Sep},
  publisher = {American Physical Society},
  doi = {10.1103/PhysRevE.110.034312},
  url = {https://link.aps.org/doi/10.1103/PhysRevE.110.034312}
}

@article{PhysRevE.106.064310,
  title = {Class of models for random hypergraphs},
  author = {Barthelemy, Marc},
  journal = {Phys. Rev. E},
  volume = {106},
  issue = {6},
  pages = {064310},
  numpages = {10},
  year = {2022},
  month = {Dec},
  publisher = {American Physical Society},
  doi = {10.1103/PhysRevE.106.064310},
  url = {https://link.aps.org/doi/10.1103/PhysRevE.106.064310}
}
\bibliographystyle{plain}

\newpage
\appendix
\section{Appendix}

    \begin{figure}[htb!]
        \centering
        \includegraphics[width=0.8\linewidth]{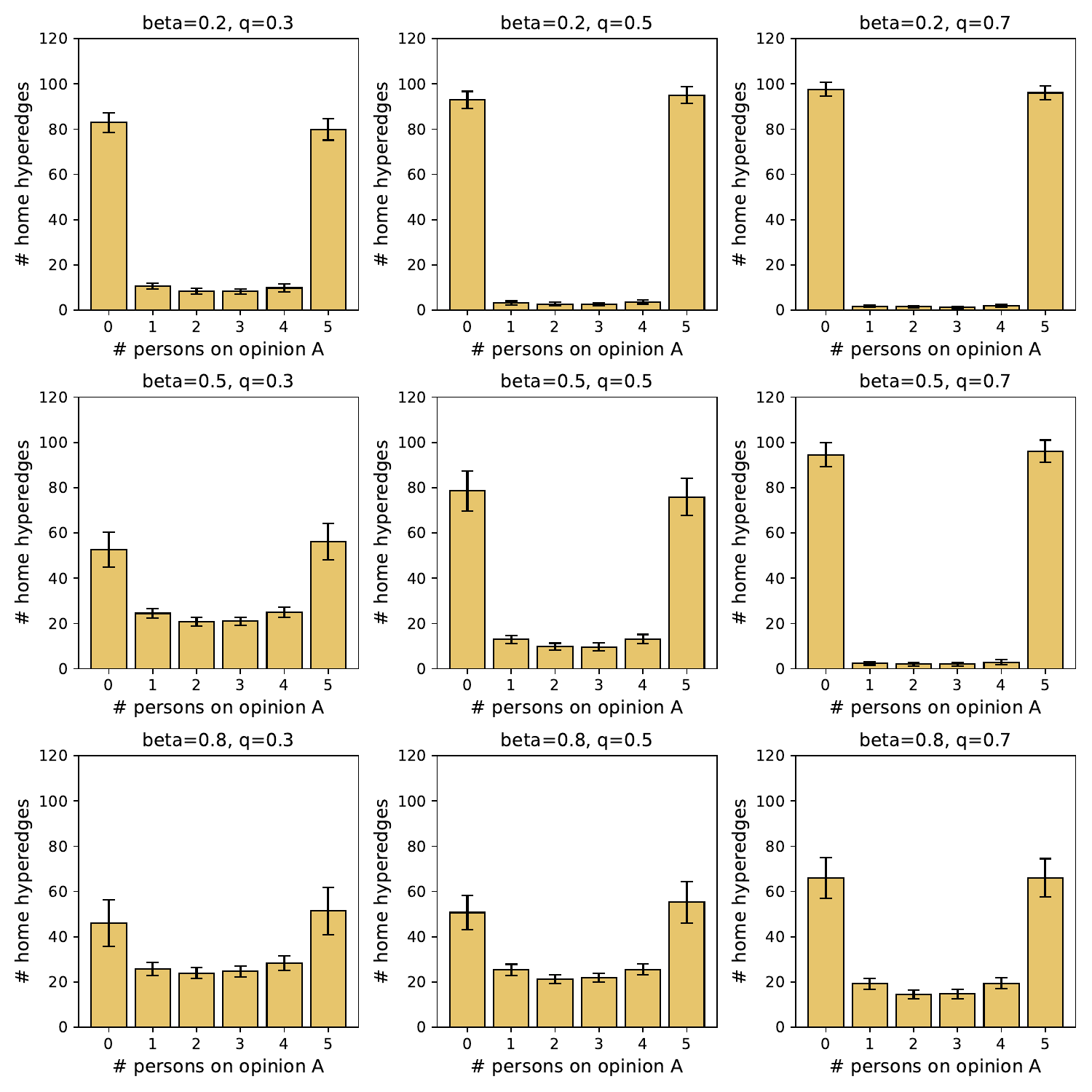}
        \caption{The number of household hyperedges with the number of opinions on "A" equal to $0,1,2,\dots, 5$ in the linear model. We see that in every case, the hyperedges with $0$ and $5$ opinions on "A" are overrepresented.}
        \label{fig:hh_op_dist}
    \end{figure}

    \begin{figure}[htb!]
        \centering
        \includegraphics[width=0.8\linewidth]{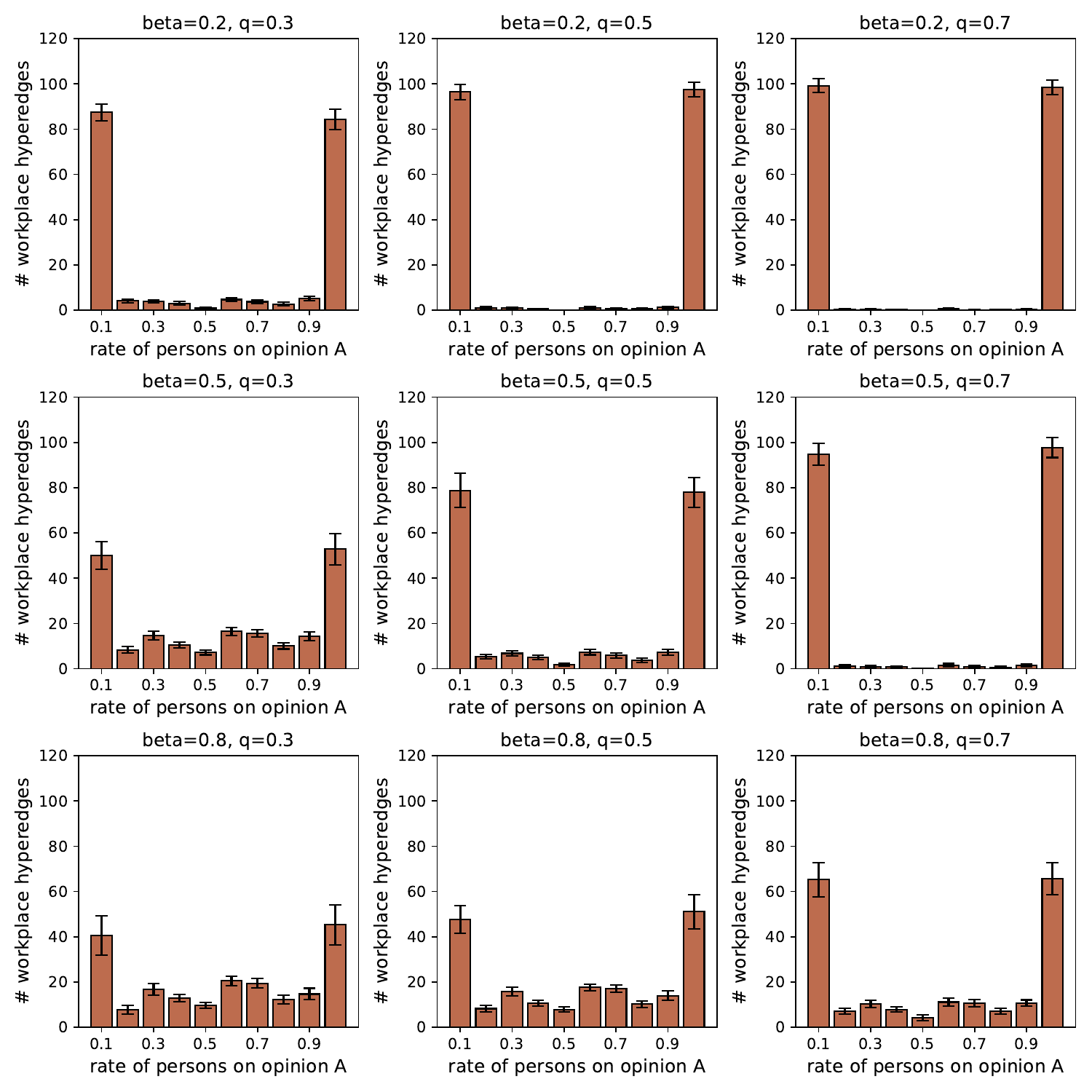}
        \caption{Distribution of the proportion of opinion A in the workplace hyperedges for different values of $\beta$ and $q$ for the linear model (notice that the second column is somewhat smaller due to error from discretization)}
        \label{fig:wp_op_dist}
    \end{figure}

    \begin{figure}[htb!]
    \centering
    \includegraphics[width=0.8\linewidth]{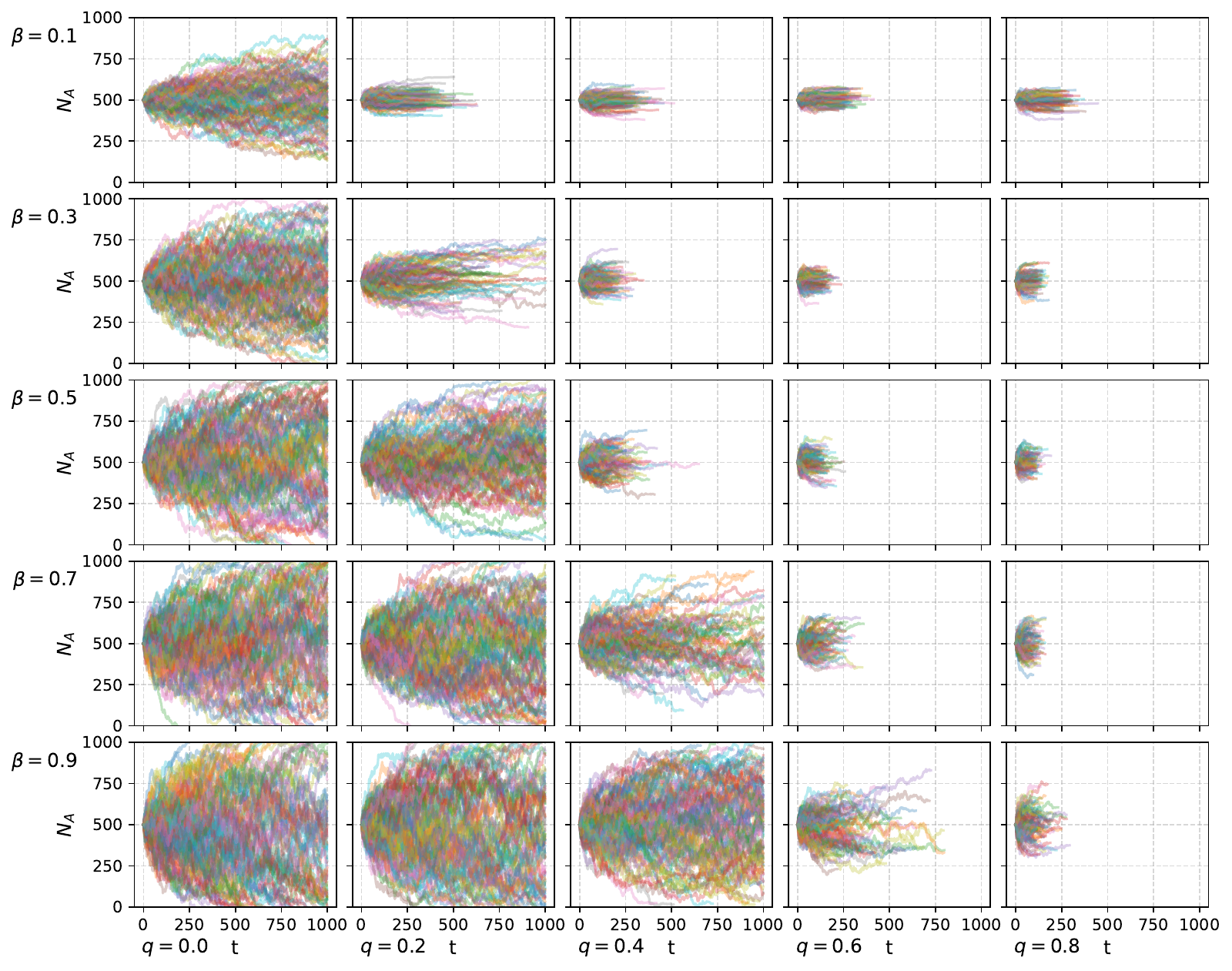}
    \caption{We ran the linear model simulations for $10^6$ steps, so every individual was chosen expectedly $1000$ times to make the iterative step in a population of $1000$. These processes started at a randomly chosen state of opinions where 500 nodes have opinion A and 500 nodes have opinion B.}
    \label{fig:sumofA_lin}
    \end{figure}

    \begin{figure}[htb!]
    \centering
    \includegraphics[width=0.8\linewidth]{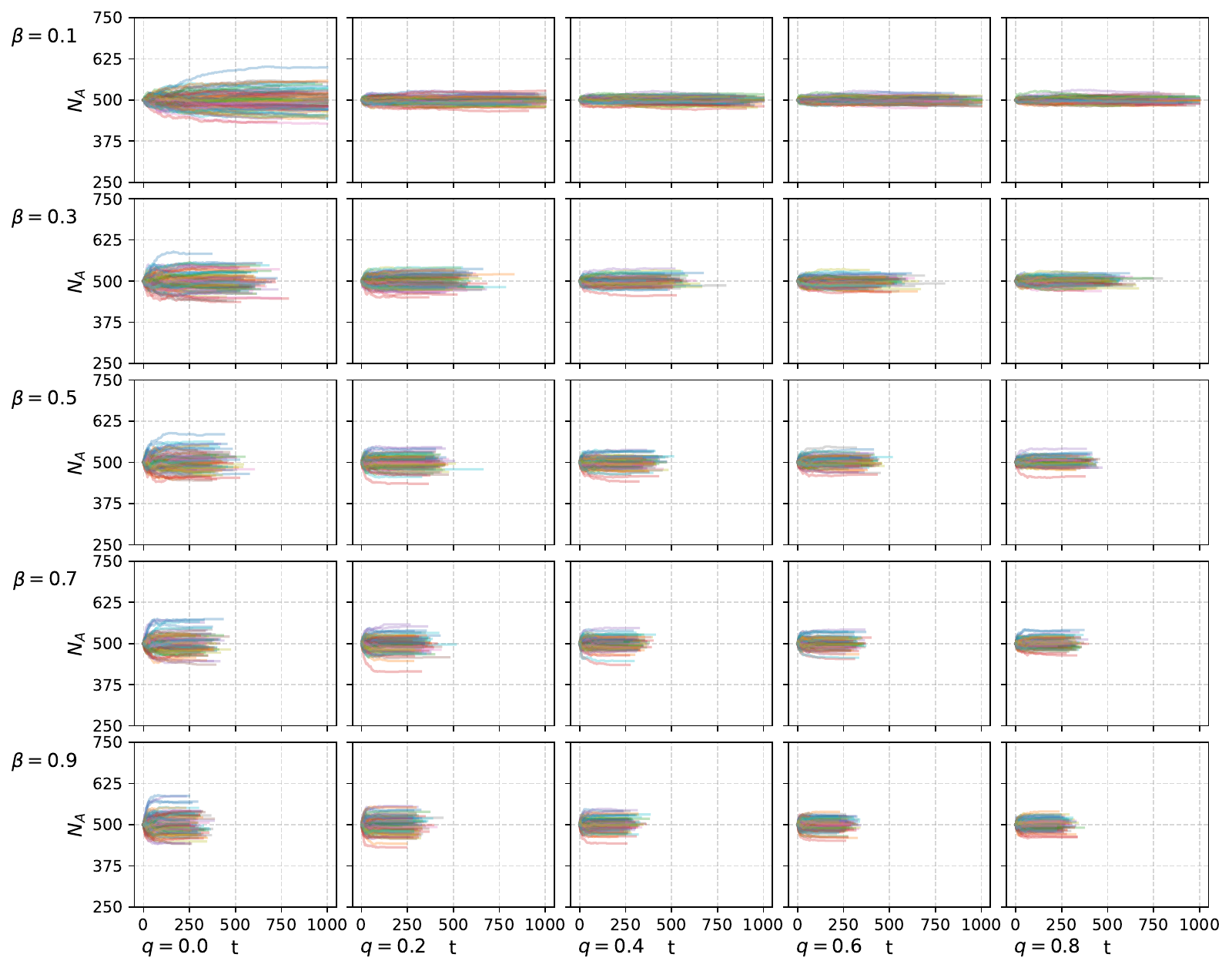}
    \caption{We ran the nonlinear model simulations for $10^6$ steps, so every individual was chosen expectedly $1000$ times to make the iterative step in a population of $1000$. These processes started at a randomly chosen state of opinions where 500 nodes have opinion A and 500 nodes have opinion B.}
    \label{fig:sumofA_nonlin}
    \end{figure}
\begin{figure}[htb!]
        \centering
        \includegraphics[width=0.8\linewidth]{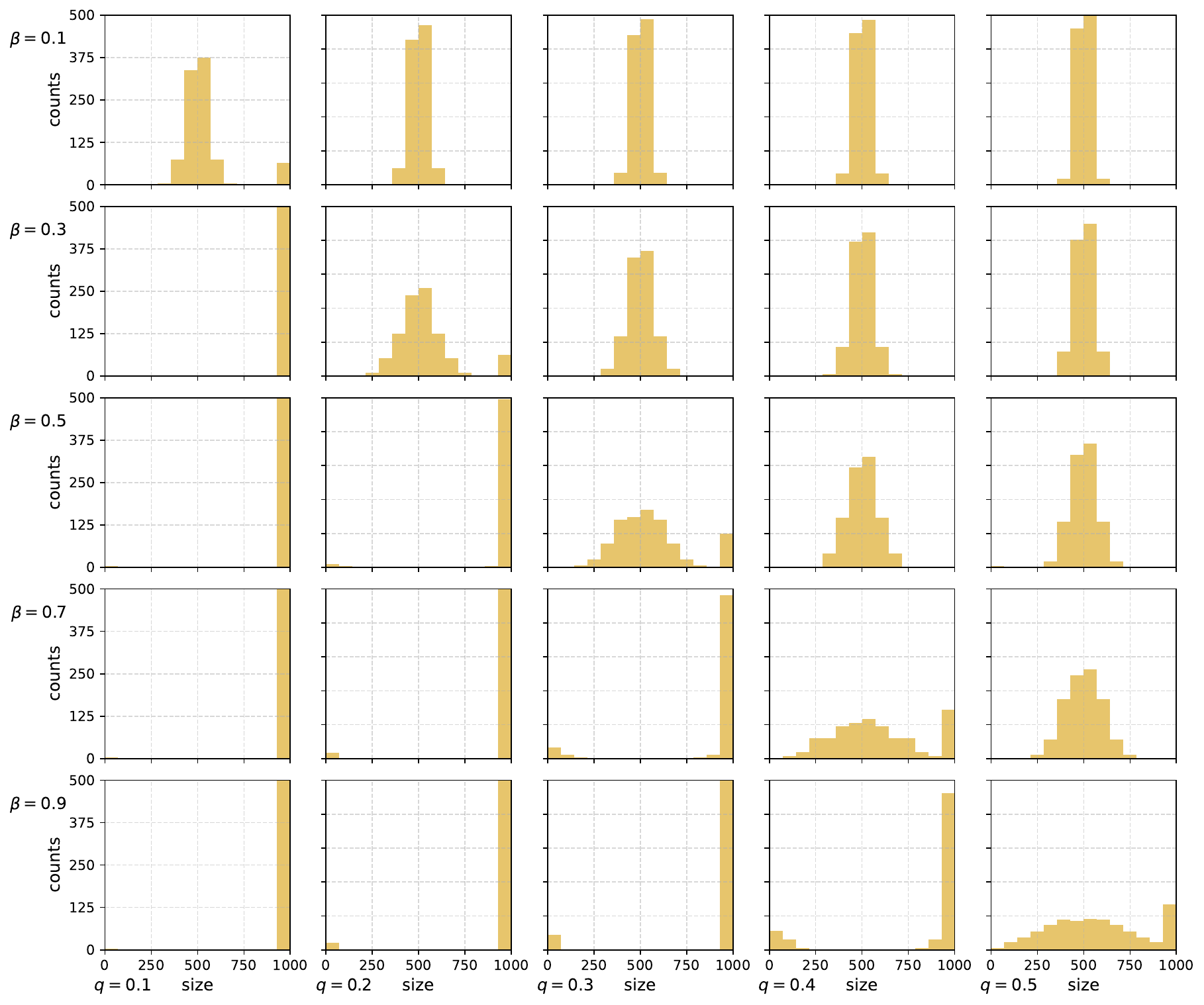}
        \caption{Component sizes}
        \label{fig:comp_sizes}
    \end{figure}
\begin{figure}[htb!]
        \centering
        \includegraphics[width=0.8\linewidth]{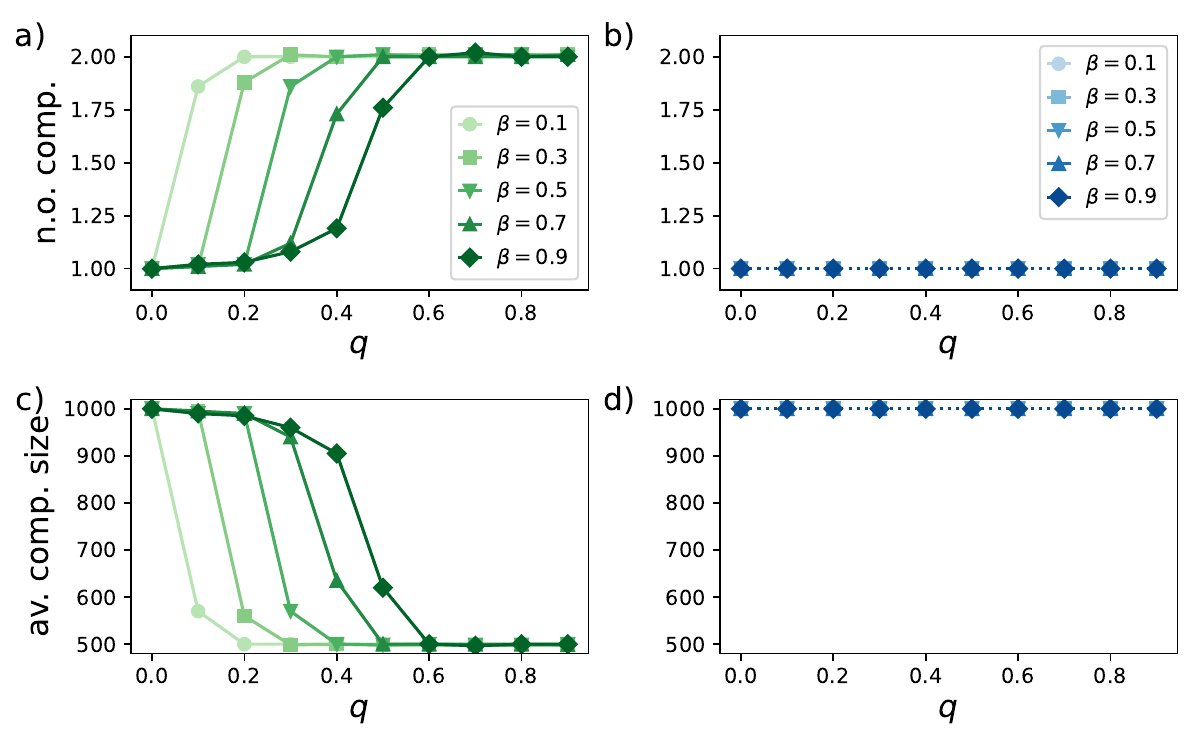}
        \caption{Number of Components}
        \label{fig:comp_num}
    \end{figure} 

\end{document}